\documentclass[11pt,a4paper,prd,superscriptaddress,nofootinbib]{revtex4}
\pdfoutput=1
\usepackage[latin1]{inputenc}
\usepackage[english]{babel}
\usepackage{amssymb}
\usepackage{amsmath}
\usepackage{amsthm}
\usepackage[]{graphicx}
\usepackage[]{subfigure}
\usepackage{tensor}
\usepackage{color}
\usepackage{cancel}
\usepackage{setspace}
\usepackage{fancyhdr}
\usepackage[bookmarks,linktocpage, colorlinks=true, plainpages=false, citecolor=blue,
                                       linkcolor=blue, urlcolor=blue, filecolor=blue ]{hyperref}

\newcommand{\smallWidthLeft}{239pt}
\newcommand{\smallWidthRight}{229pt}
\newcommand{\thirdWidthLeft}{159.33pt}
\newcommand{\thirdWidthRight}{149.33pt}
\newcommand{\fullWidth}{348pt}

\newcommand{\bea}{\begin{eqnarray}}
\newcommand{\eea}{\end{eqnarray}}
\newcommand{\be}{\begin{equation}}
\newcommand{\ee}{\end{equation}}
\newcommand{\beast}{\begin{eqnarray*}}
\newcommand{\eeast}{\end{eqnarray*}}
\newcommand{\pkt}{\; .}
\newcommand{\kma}{\; ,}

\def\e{{\rm e}}

\begin{document}

\title{Negative Modes of Oscillating Instantons}

\author{Lorenzo Battarra}
\email{lorenzo.battarra@aei.mpg.de }
\affiliation{Max-Planck-Institut f\"ur Gravitationsphysik, Albert-Einstein-Institut \\
Am M\"uhlenberg 1, D-14476 Potsdam, Germany}

\author{George Lavrelashvili }
\email{lavrela@itp.unibe.ch }
\affiliation{Department of Theoretical Physics,
A.Razmadze Mathematical Institute \\
I.Javakhishvili Tbilisi State University,
GE-0177 Tbilisi, Georgia}

\author{Jean-Luc Lehners $^1$}
\email{jean-luc.lehners@aei.mpg.de}

\date{\today}

\begin{abstract}
We investigate in detail the properties of oscillating instanton solutions discussed recently
in the literature. We find that the solutions with $N$ nodes contain exactly $N$ homogeneous
negative modes in their spectrum of linear perturbations.
The existence of extra negative modes for the $N>1$ solutions suggest that they
are not final state physical objects resulting from tunneling,
but rather unstable intermediate thermal configurations. By contrast, the single negative mode for the $N=1$ instanton confirms its interpretation as mediating the curved-space tunneling between vacua with equal energy densities.
\end{abstract}

\maketitle

\section{Introduction}
Modern developments showed \cite{Bousso:2000xa,Douglas:2003um} that string theory predicts the existence of a multitude of vacua,
some of which being stable and some metastable. If this picture turns out to be correct, then, via the mechanism of eternal inflation (see the review \cite{Guth:2007ng} and references therein), it will have profound consequences for the ultra-large-scale structure of the Universe.

The theory of metastable vacuum decay in flat spacetime was developed long
ago \cite{Langer:1967ax,Langer:1969bc,Kobzarev:1974cp,Coleman:1977py,Callan:1977pt,Coleman:1978ae}.
It was shown that metastable
vacuum decay proceeds via the nucleation of bubbles of true vacuum within the false vacuum and the subsequent growth of these bubbles.
Within the Euclidean approach the bubble nucleation process is described by the so-called "bounce" \cite{Coleman:1977py},
which refers to a classical solution of the Euclidean equations of motion with certain boundary conditions.
It was shown that in the WKB approximation the action of the bounce determines the tunneling rate exponent.
Furthermore, in flat space-time the pre-exponential factor in the decay rate was calculated
by taking into account quadratic fluctuations about the classical solution \cite{Callan:1977pt,Isidori:2001bm,Baacke:2003uw,Dunne:2005rt}.
Its value is given by a ratio of the functional determinants of
the fluctuation operators corresponding to the bounce and the metastable vacuum. It is important to note that there is exactly one negative mode
in the spectrum of small perturbations about the bounce solution in flat space-time \cite{Callan:1977pt,Maziashvili:2002jb}.
This single negative mode is essential in making the decay picture coherent \cite{Coleman:1987rm}.

False vacuum decay with gravity was first investigated by Coleman and De Luccia \cite{Coleman:1980aw}.
It was shown that, as in a flat space-time, the Euclidean action of the bounce
solution determines the leading exponent in the false vacuum decay rate and that the analytic continuation of the bounce
defines the space-time geometry at the moment of true vacuum bubble materialization.
The Coleman-De Luccia results were reconsidered and justified by using
a technique that explicitly accounts for the structure of an initial
state of quantum field in the semiclassical calculations of the path integral in
curved space-time \cite{Rubakov:1999ir} and more recently by a thermal derivation \cite{Brown:2007sd}.
In addition to Coleman-De Luccia bounces, excited multi-bounce solutions have been discussed in the literature
\cite{Bousso:1998ed,Hackworth:2004xb}. It was suggested that in certain regimes these "oscillating" bounces most
likely will play a significant role in tunneling processes.
Instanton solutions mediating tunneling between degenerate vacua in curved space were
investigated in \cite{Hackworth:2004xb,Lee:2009bp} and excitations thereof were studied in \cite{Lee:2011ms}.
Possible observational tests of cosmological instantons were considered in \cite{Gratton:1999hv}.
Gravitational corrections to standard model vacuum decay were studied in \cite{Isidori:2007vm}
and implications of recent ATLAS and CMS experimental results about the Higgs boson mass
for the stability of the electroweak vacuum were investigated in \cite{EliasMiro:2011aa}.
In spite of much work which has been done since Coleman and De Luccia's paper
on taking into account gravity in tunneling processes, there still remain many open questions,
{\it cf.} discussions in \cite{Banks:2002nm,Bajc:2011iu,Garriga:2012qp}.

While in flat space-time finding a negative mode about a bounce is a straightforward task, when gravity is taken into
account it becomes a more involved problem \cite{Lavrelashvili:1985vn,Tanaka:1992zw,Lavrelashvili:1998dt,
Khvedelidze:2000cp,Lavrelashvili:1999sr,Gratton:2000fj,Lavrelashvili:2006cv,Dunne:2006bt}.
Using Dirac's theory of constrained Hamiltonian systems, it was shown that with the proper reduction procedure
one finds a single negative mode about Coleman-De Luccia bounces \cite{Khvedelidze:2000cp,Lavrelashvili:1999sr}.
Furthermore it was demonstrated in \cite{Lavrelashvili:2006cv} that the excited multi-bounce solutions possess more then one
negative mode and consequently they do not contribute directly to the tunneling processes, but rather specify a
path through configuration space that connects thermally excited horizon volume configurations
(for a discussion of this point see \cite{Brown:2007sd,Brown:2011um}).

In contrast to bounces, instanton solutions in flat spacetime typically only have zero modes, and no negative modes. As such, they describe the quantum mixing between equal energy states of the system
(see {\it e.g.} the book \cite{rajaraman}), rather than describing the decay of one vacuum into another vacuum of equal energy density.
The aim of the present investigation is to better understand what happens when gravity is included. For this purpose, we will study in detail the  properties of the oscillating instanton solutions discussed in the literature recently
\cite{Lee:2009bp,Lee:2011ms}. As we will show, in the presence of gravity the standard instantons, which interpolate monotonically between one vacuum and the other, possess exactly one negative mode, and they do indeed describe a tunneling process between degenerate vacua. By contrast, oscillating instantons contain additional negative modes which invalidate their interpretation as contributing to a decay process. Along the way, we will clarify a number of issues that arise in the limit where the minima of the potential become degenerate in height, such as the proper definition of the thin-wall approximation and the interpretation of the existence of the bubble wall itself.

The rest of paper is organized as follows: in the next Sect. we recall the main properties of instantons and
bounces using simple quantum mechanical examples. In Sect. III we review details of the process of tunneling with gravity
and we also review gravitational instanton solutions and their properties.
In Sect. IV we present our analysis of linear perturbations and their spectrum about instanton solutions.
Sect. V contains concluding remarks.

\section{Euclidean Solutions in Flat Space-Time: Instantons vs Bounces}

\begin{figure}[t]
\begin{minipage}{\smallWidthLeft}
\includegraphics[width=\smallWidthRight]{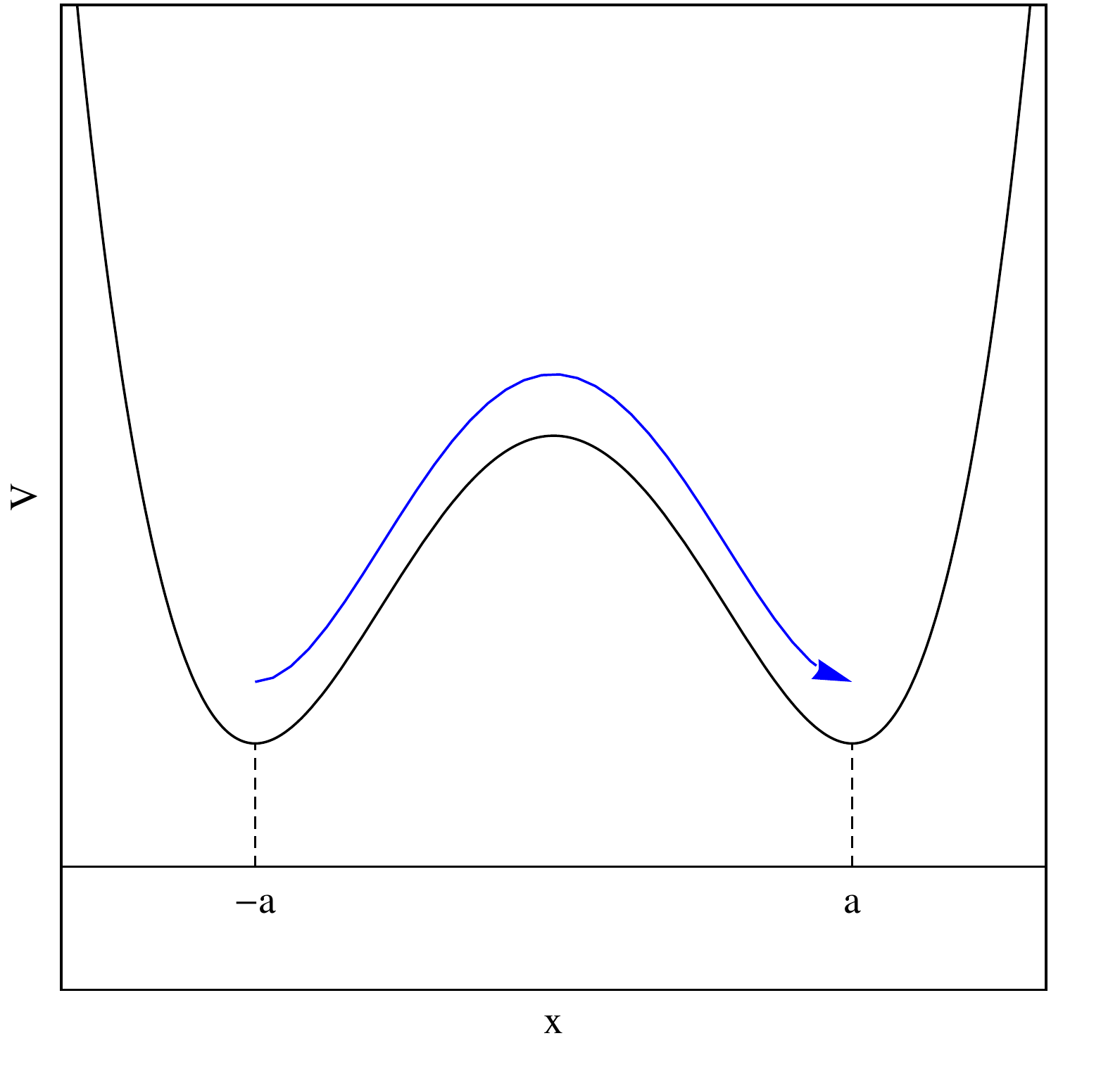}
\end{minipage}%
\begin{minipage}{\smallWidthRight}
\includegraphics[width=\smallWidthRight]{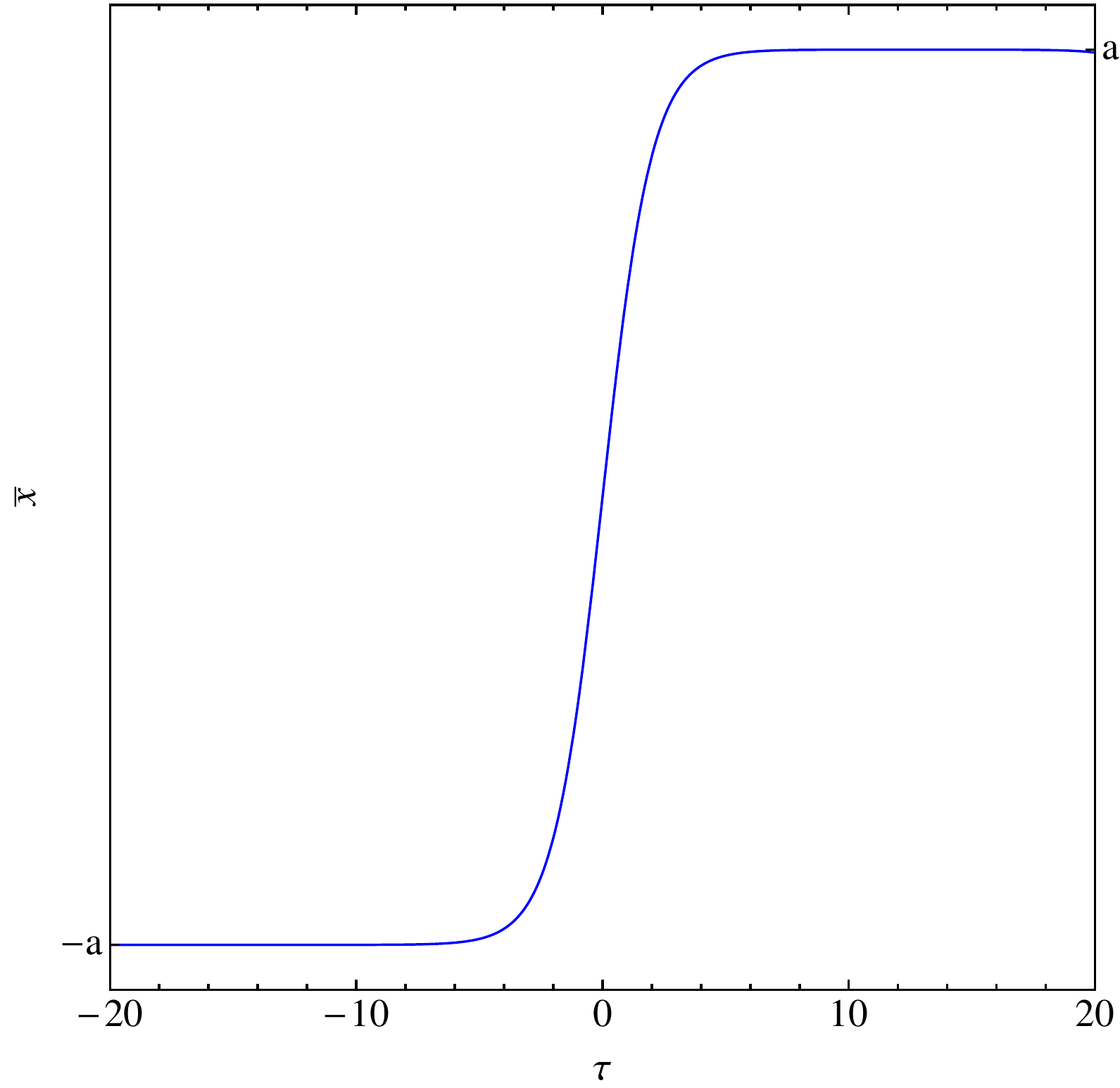}
\end{minipage}\caption{\small \label{fig:QM} The left panel exhibits the shape of a double well
potential with two degenerate minima located at $ x = \pm a$. The blue curve represents the instanton solution, whose profile is depicted in the right panel.}
\end{figure}

Let us consider the basic example of one dimensional field theory (quantum mechanics)
and assume that the potential $V(x)$ has a double well shape with the degenerate vacua at $x=\pm a$, see Figure \ref{fig:QM}.
Classically, the ground state of a particle at rest in one of the minima is a stable state.
Quantum mechanically it is well known that the existence of instantons,
{\it i.e.} finite action Euclidean solutions interpolating between $-a$ and $a$ ,
leads to a splitting of the ground state energy
(see {\it e.g.} the detailed discussion in \cite{Coleman:1978ae},
which we briefly follow in the present Sect.).

In the semiclassical approximation, summing multi-instanton
configurations   one finds for the two lowest-lying energy eigenstates
\be
E_\pm =\frac{\hbar \omega}{2} \pm \hbar K \e^{-S/\hbar} [1+O(\hbar)] \kma
\ee
where the exponent $S=S[\bar{x}]$ is the Euclidean action of the instanton solution
and the pre-exponential factor
\be \label{eq:K_inst}
K=\frac{1}{2} \left(\frac{S}{2\pi\hbar} \right)^{1/2}
\left(\frac{ \textrm{det}'[-\partial_t^2+V''(\bar x) ]}
{ \textrm{det}[-\partial_t^2+\omega^2]} \right)^{-1/2}
\ee
is calculated by taking the Gaussian integral over the quadratic action
describing linear perturbations about the instanton.
There is one translational zero mode in the spectrum
which needs special care and
$\rm{det'}$ means that this zero mode is treated separately.
Integration over it gives the normalization factor in \eqref{eq:K_inst}. Thus, tunneling effects remove the degeneracy and shift the ground state energy. However, the correction to the energy levels is real, and two classically stable states, localized around either minimum, are promoted to two non--localized,  stationary quantum states.

When the minima of the potential have different energy, the semi--classical analysis yields qualitatively different results. Now there is one vacuum with higher energy, which is
a false (metastable) vacuum, and another with lower energy which constitutes the true (stable) vacuum. In this case, the bounce solution describes the decay of the false vacuum which proceeds via the nucleation of true vacuum bubbles in the false vacuum.
In the semi-classical approximation, summing the contributions of multi-bounce
configurations one finds the following correction to the energy of the false vacuum
\be
E=\frac{\hbar \omega}{2}-\hbar K \e^{-S/\hbar} [1+O(\hbar)] \kma
\ee
where
\be \label{eq:K_bounce}
K=\frac{1}{2}  \left( \frac{S}{2\pi\hbar} \right)^{1/2}
 \left(\frac{ \textrm{det}'[-\partial_t^2+V''(\bar x) ]}
{ \textrm{det}[-\partial_t^2+\omega^2]} \right)^{-1/2} \pkt
\ee
It is remarkable that there exists exactly one negative mode in the spectrum of linear perturbations
about the bounce solution. This can be inferred from the fact that the (translational) zero energy wave function
${ \psi_0\sim \frac{d\bar x}{dt}}$  of the corresponding Schr\"odinger equation has a node.
The negative mode implies that the correction \eqref{eq:K_bounce} to the false vacuum energy  is purely imaginary
and thus we actually have a decay process. So, the false vacuum decay probability per unit time is given by
\bea
{\Gamma }= { -2 \, {\rm Im} E/\hbar}
=  \left( \frac{S}{2\pi \hbar} \right)^{1/2}
 \left|\frac{det'[-\partial_t^2+V''(\bar x) ]}
{det[-\partial_t^2+\omega^2]} \right|^{-1/2}~~\e^{-S/\hbar}~~[1+O(\hbar)] \pkt
\eea
Analyzing metastable vacuum decay processes in
the 1988 NPB article ``Quantum Tunneling and Negative Eigenvalues'' \cite{Coleman:1987rm}
Coleman arrives at the strong conclusion:
``There may exist solutions in other ways like bounces and
which have more than one negative eigenvalue, but, even if they do exist,
they have nothing to do with tunneling."
So, determining the number of negative modes in the spectrum of perturbations is of great importance in finding the proper physical interpretation of any given solution.

\section{Tunneling with Gravity}

\begin{figure}[t]
\begin{minipage}{\smallWidthLeft}
\includegraphics[width=\smallWidthRight]{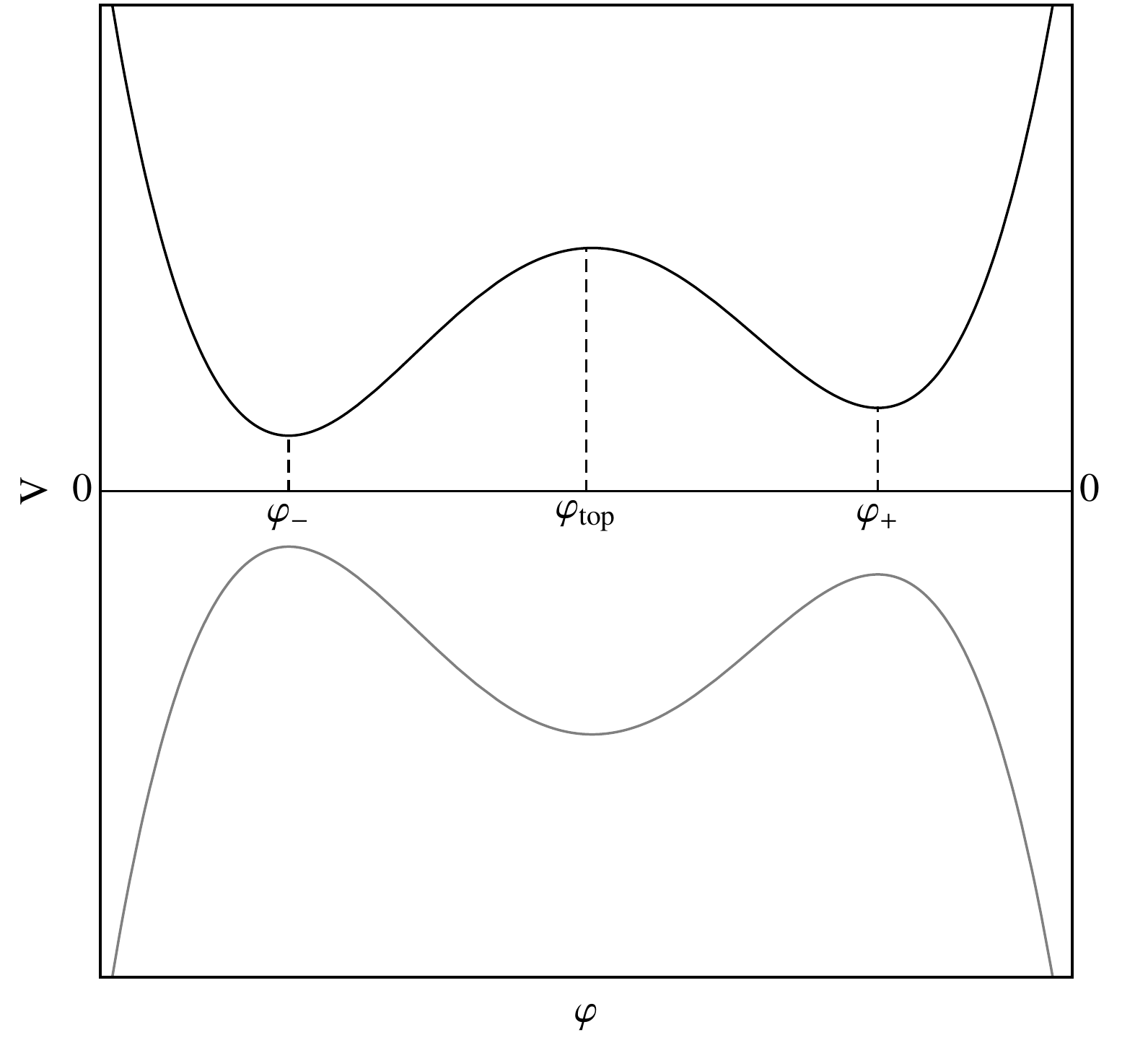}
\end{minipage}%
\begin{minipage}{\smallWidthRight}
\includegraphics[width=\smallWidthRight]{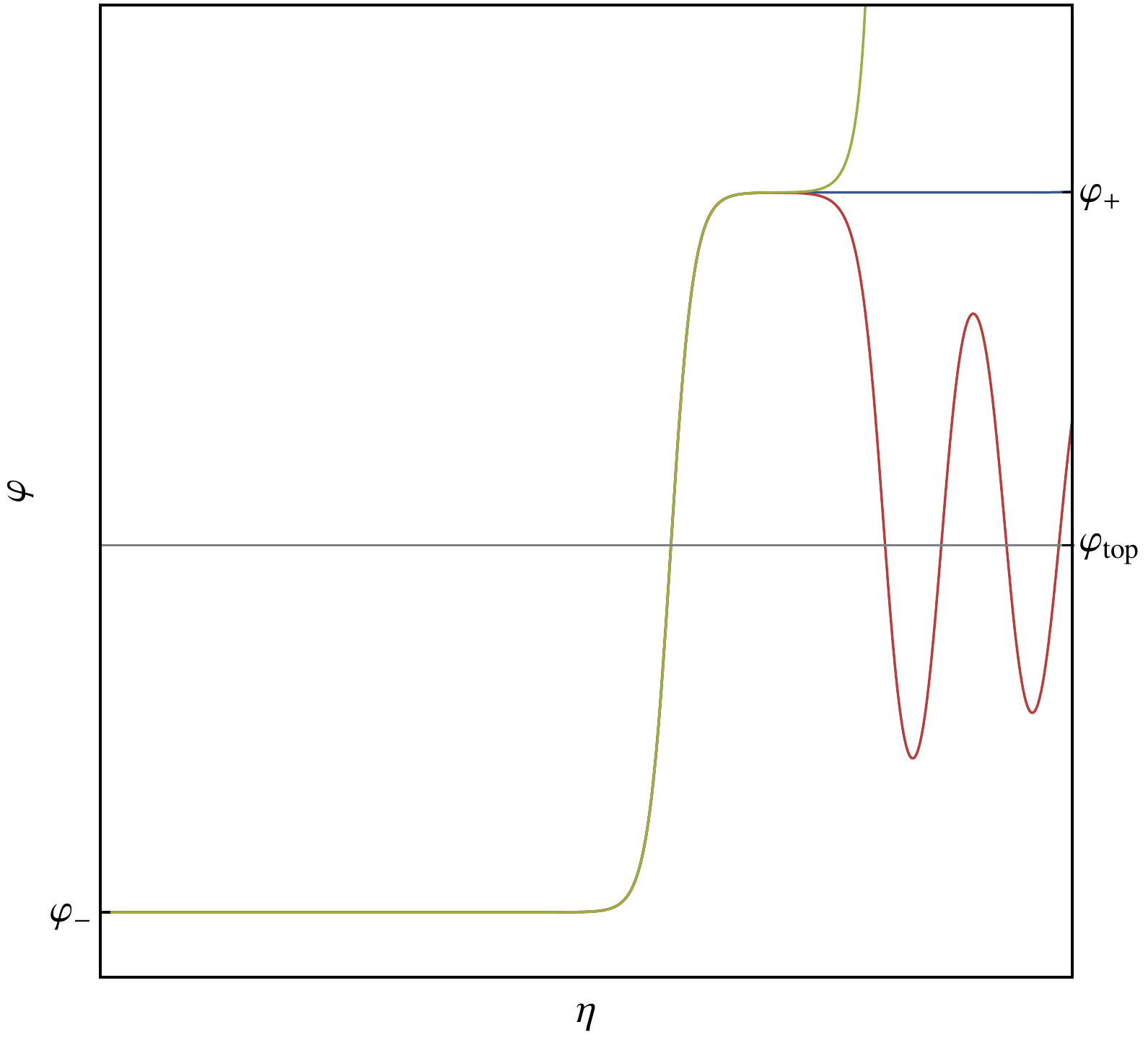}
\end{minipage}\caption{\small \label{fig:generalPotential} The left panel shows the shape of a double well
potential with two non--degenerate vacua. The gray curve represents the inverse potential,
driving the solution of the Euclidean field equations. In the right panel,
the flat space bounce appears as the separating solution between undershooting (red curve)
and overshooting (green curve).}
\end{figure}

Let us now consider a self-interacting scalar field theory minimally coupled
to Einstein gravity in four dimensions:
\begin{equation}
S = \int d ^4x \sqrt{-g} \, \left( \frac{1}{2 \kappa} R
- \frac{1}{2} \nabla_\mu \varphi \nabla^\mu \varphi - V( \varphi) \right) \kma
\end{equation}
where $\kappa = 8 \pi G$ is the reduced Newton's constant.
We consider potentials $V(\varphi)$ which are bounded from below and admit two (possibly degenerate) vacua
at $\varphi = \varphi _{\pm}$, separated by a barrier whose top is located at $\varphi = \varphi_{top}$
(see Figure \ref{fig:generalPotential}). We denote $V _{ \pm} \equiv V( \varphi _{\pm})$ and conventionally choose
$V _{-} \leq V _{+}$. In other words, when the vacua are non--degenerate, $ \varphi _{-}$ labels
the position of the true vacuum in field space.

\subsection{Field Equations}

In the semi--classical limit, the rate for the tunneling process is again described by solutions of the Euclidean field
equations with proper boundary conditions.
In flat spacetime it was shown \cite{Coleman:1977th} that the $O(4)$--invariant solution has least Euclidean action,
and provides the dominant contribution to the decay process. An analogous proof does not exist in the case where gravity is included -- rather the $O(4)$ invariance is still a conjecture \cite{Coleman:1980aw} that we will assume to be true. Thus, the Euclidean metric and scalar field Ans\"atze can be written as
\begin{eqnarray}
ds^2 & = & d \eta ^2 + \rho^2( \eta) d \Omega_3^2 \;, \\
\varphi & = & \varphi ( \eta) \;.
\end{eqnarray}
The field equations read
\begin{eqnarray} \label{eq:curvature}
\rho ^{\prime \prime} & = & - \frac{ \kappa \rho}{3} \left( \varphi ^{ \prime 2} + V \right) \;,\\ \label{eq:fieldEq}
\varphi '' & = & V_{, \varphi} - 3 \frac{ \rho^\prime}{ \rho} \varphi ' \;, \\ \label{eq:constraint}
 \rho ^{ \prime 2} & = & 1 + \frac{ \kappa \rho ^2}{3} \left( \frac{1}{2} \varphi ^{ \prime 2} - V \right) \kma
\end{eqnarray}
where a prime denotes differentiation w.r.t. Euclidean proper time
$\eta$ and $V_{, \varphi} \equiv \frac{dV}{d\varphi}$. Taking the origin $ \rho = 0$ to be conventionally located at $ \eta = 0$, regularity imposes the following conditions:
\begin{eqnarray} \label{eq:bc1}
\rho'( \eta = 0) & = & 1 \;,
\\ \label{eq:bc2}
\varphi'( \eta = 0) & = & 0 \;.
\end{eqnarray}
Together with \eqref{eq:fieldEq}, these equations can be thought of as describing the one-dimensional motion of a unit
mass particle which starts at rest at $ \eta = 0$ and is subject to an inverted potential $ -V( \varphi)$. According to the
sign of $ \rho'/ \rho$, a friction/anti--friction term is also present, which depends non--linearly on the motion
of the particle via \eqref{eq:constraint}.

When $ \varphi _0 \equiv \varphi( \eta = 0)$ is a stationary point of $V$, the corresponding solutions are maximally
symmetric with a constant scalar field profile $ \varphi = \varphi_0$\footnote{By taking derivatives of \eqref{eq:fieldEq}
one can easily see that all the derivatives of $ \varphi$ vanish at $ \eta = 0$.}.
Depending on the sign of $V_0 \equiv V( \varphi _0),$ one obtains
\begin{eqnarray} \label{eq:dSGeometry}
V_0 > 0: \quad & \textrm{Euclidean dS (}S^4 \textrm{)} & \quad\rho = H ^{-1} \sin{ \left(H \eta \right)},
\quad H ^2 \equiv \frac{ \kappa V_{0}}{3} \;, \\
V_0 = 0: \quad& \textrm{Euclidean space (}E_4 \textrm{)} & \quad\rho = \eta \;, \\
V_0 < 0: \label{eq:AdSGeometry}\quad& \textrm{Euclidean AdS} & \quad \rho = \sqrt{ \frac{ 3}{ \kappa |V_0|}}
\sinh{ \left(\sqrt{ \frac{ \kappa |V_0|}{3}} \eta \right)} \;.
\end{eqnarray}
Solutions with a non--trivial profile for the scalar field and which interpolate between the true vacuum
and the false vacuum, {\it i.e.} for which we impose the boundary conditions
\begin{eqnarray*}
\varphi_0 \equiv \varphi( \eta = 0) & \simeq & \varphi _{ \pm} \;, \\
\bar{ \varphi} \equiv \varphi( \eta = \bar{ \eta}) & \simeq & \varphi _{ \mp} \;,
\end{eqnarray*}
also exist. If they contain a single negative eigenmode in their fluctuation spectrum, they describe the decay of one vacuum into the other. The value of $ \bar{ \eta}$ can be either finite, as for the
maximally symmetric $S^4$ solution:
\begin{equation}
\bar{ \eta} = \pi H ^{-1} \;,
\end{equation}
or infinite, as in the other cases $ V _0 \leq 0$. We refer to these two types of solutions as \textit{compact}
and \textit{non--compact} bounces respectively.

\subsection{Bounces Connecting dS Vacua \label{sect:bouncesdS}}

Let us focus on the case where both vacua are of the de Sitter type, and the potential is bounded by a positive constant,
as in Figure \ref{fig:generalPotential}.
In \cite{Hackworth:2004xb} it was shown that the corresponding theories can admit, depending on the details of the
potential, several compact bounce solutions. These can be labeled by the number of scalar field oscillations between the two vacua.
Relying on analytic arguments, the number of such solutions was determined to be equal to the largest integer $N$ such
that (see also \cite{Jensen:1983ac})
\begin{equation} \label{eq:weinbergNumber}
N(N+3) < \frac{|V _{, \varphi \varphi}( \varphi_{top})|}{H_{top}^2}, \quad H_{top} ^2
\equiv \frac{ \kappa V( \varphi_{top})}{3} \;.
\end{equation}
We may obtain a qualitative understanding of the existence of such solutions by analyzing the field equations.
First, one can easily prove that, $V$ being bounded from below by a positive constant, all solutions to the field
equations that are regular at $ \eta = 0$ (see \eqref{eq:bc1} and \eqref{eq:bc2}) are compact. Indeed:
\begin{eqnarray} \label{eq:restoringForce}
\rho ^{\prime \prime}  & = &  - \frac{ \kappa \rho}{3} \left( \varphi ^{ \prime 2} + V \right) \leq - H_{min}^2\,
\rho, \quad  H_{min}^2 \equiv \frac{k}{3} \textrm{min}(V) > 0 \;, \\ \label{eq:restoringForceBcs}
\rho(0) & = & 0\;.
\end{eqnarray}
One can view \eqref{eq:restoringForce} as the equation of motion of an oscillator with a time--dependent frequency,
bounded from below by a positive constant. It is easy to show that, given the initial conditions
\eqref{eq:restoringForceBcs}, $ \rho$ vanishes before the timescale set by $ H_{min}$:
\begin{equation}
\rho( \bar{ \eta}) = 0: \quad \bar{ \eta} \leq \pi H_{min}^{-1}.
\end{equation}
However, for a generic value of $ \varphi_0$, one obtains a singular solution. Indeed, when $ \eta$ approaches
$ \bar{ \eta}$, the coefficient $ \rho'/ \rho$ in the scalar field equation becomes large and negative.
For this reason, except for a discrete set of values for $ \varphi_0$, the scalar field diverges (positive or negative)
when $ \eta$ approaches $ \bar{ \eta}$. If $ \varphi_0$ equals one of those special values, the scalar field approaches
a constant $ \bar{ \varphi}_0$ when $ \eta \rightarrow \bar{ \eta}$.

Now, let us try to understand how these particular values of $ \varphi_0$ emerge from the field equations.
Like in the flat space case, setting $ \varphi_0$ arbitrarily close to the true vacuum $ \varphi_-$, one can make sure
that the corresponding solution for $ \varphi$ overshoots monotonically. However, this does not require the two
vacua to be non--degenerate anymore. Indeed, if $ \varphi_0$ is  sufficiently close to $ \varphi_-$, the coefficient
$ \rho'/ \rho$ will start taking large negative values at $ \eta \simeq \pi H _{-} ^{-1}$, when the scalar field still
sits near the true vacuum. The combined action of the potential and the anti--friction term will then kick the scalar
field out of the true vacuum and make it reach infinity at $ \eta = \bar{ \eta}$. When $ \varphi_0 - \varphi _{-}$ is
increased, the scalar field ``particle'' may lose, during the phase in which $ \rho'/ \rho$ is positive,
enough energy to be trapped and oscillate for a while around the minimum of $-V$ located at $ \varphi = \varphi _{top}$,
before being kicked out again by the anti--friction at $ \eta \sim H^{-1}$. The period of such oscillations is roughly
equal to the inverse curvature of the potential at $ \varphi_{top}$. The maximum number of oscillations is
approximately set by the ratio between these two periods:
\begin{equation}
N \lesssim \frac{|V _{, \varphi \varphi}( \varphi_{top})|^{1/2}}{H}\quad (\textrm{when }N \gg 1) \;.
\end{equation}
Regular oscillating bounce solutions appear as the separating elements between classes of solutions which have
a different number of oscillations before the overshooting: this rough estimate is a qualitative version of the
rigorous result \eqref{eq:weinbergNumber}.

\begin{figure}[thbp] \centering
\includegraphics[width=\fullWidth]{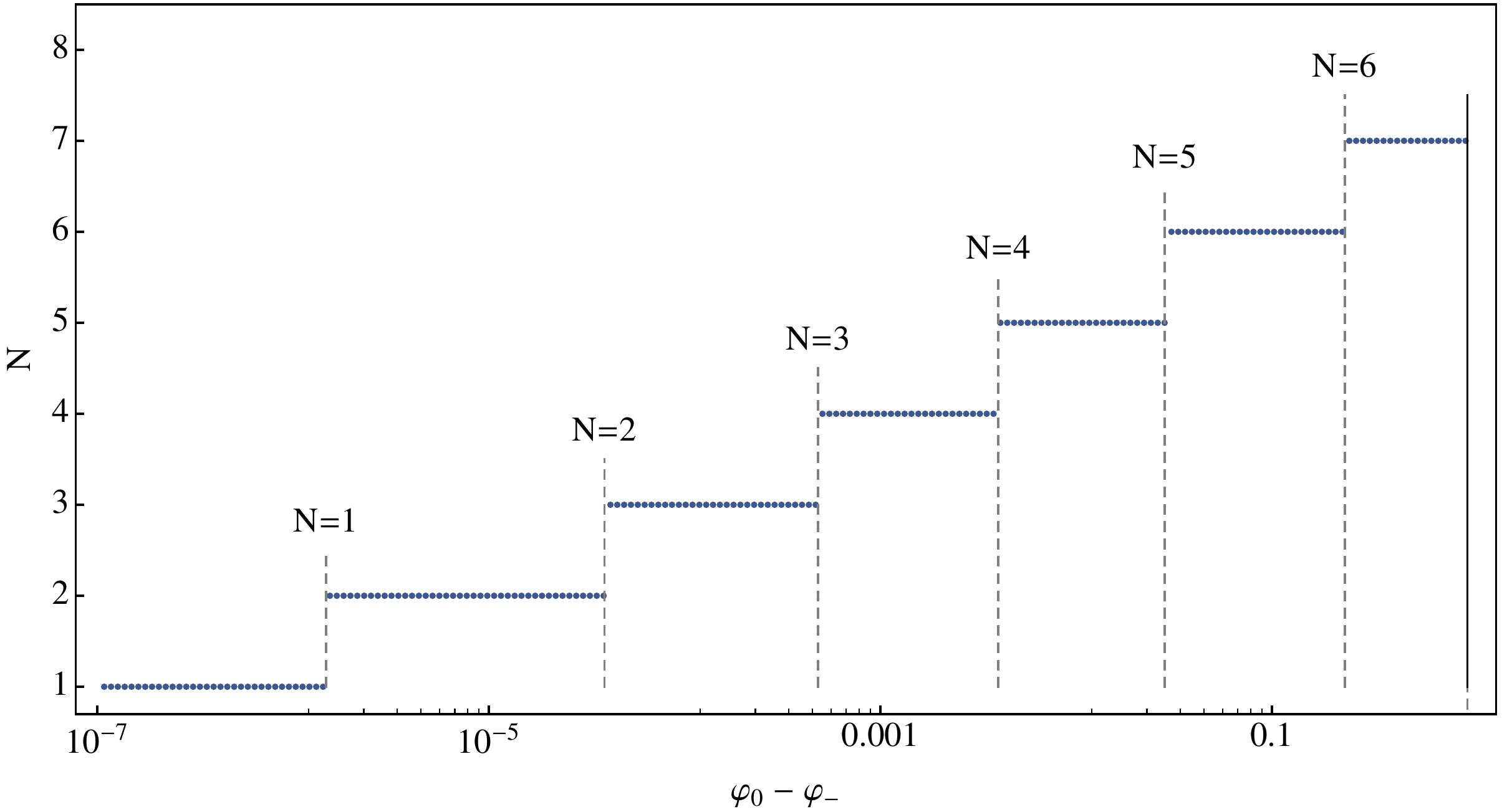} \vspace{0.4cm}\\
\begin{minipage}{\smallWidthLeft}
\includegraphics[width=\smallWidthRight]{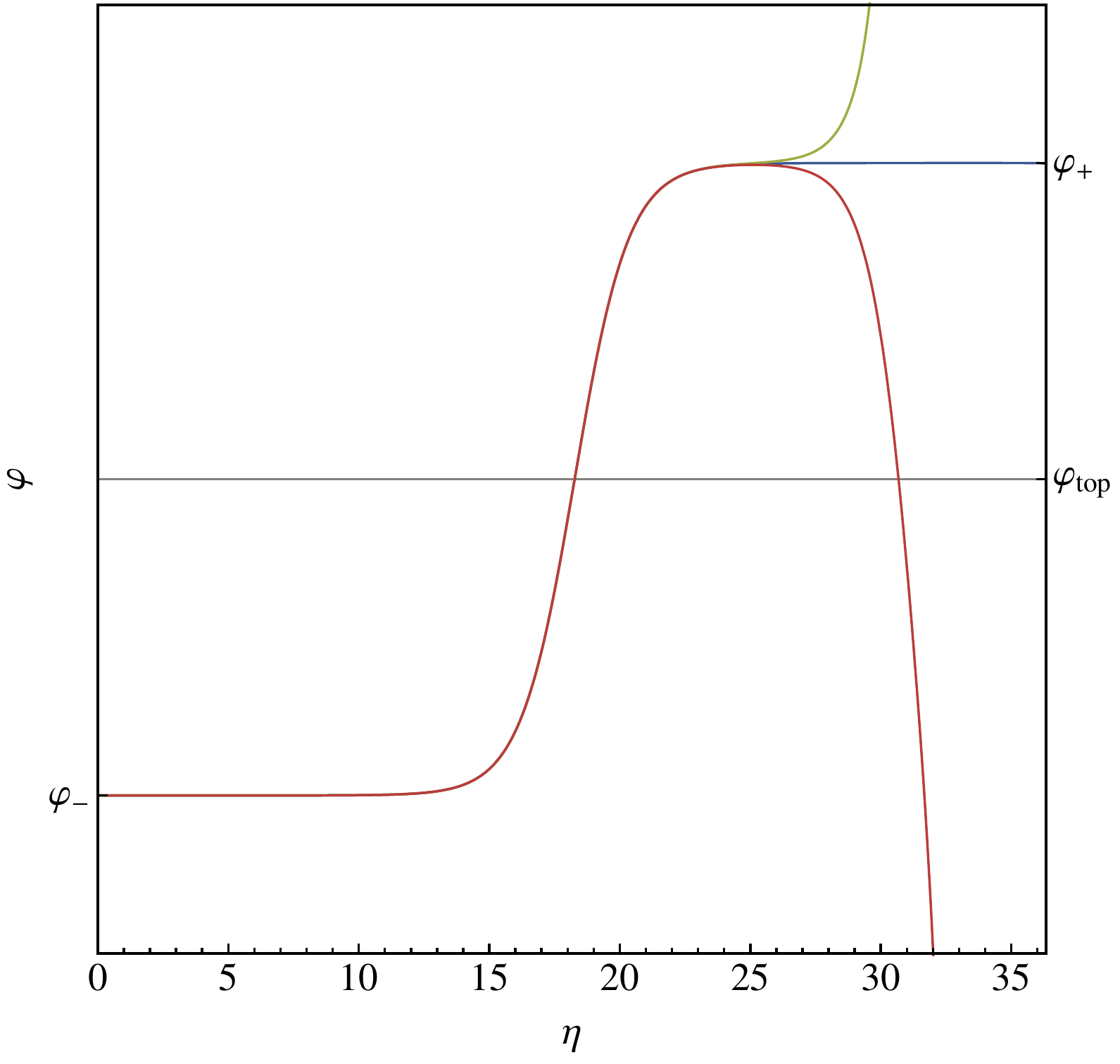}
\end{minipage}%
\begin{minipage}{\smallWidthRight}
\includegraphics[width=\smallWidthRight]{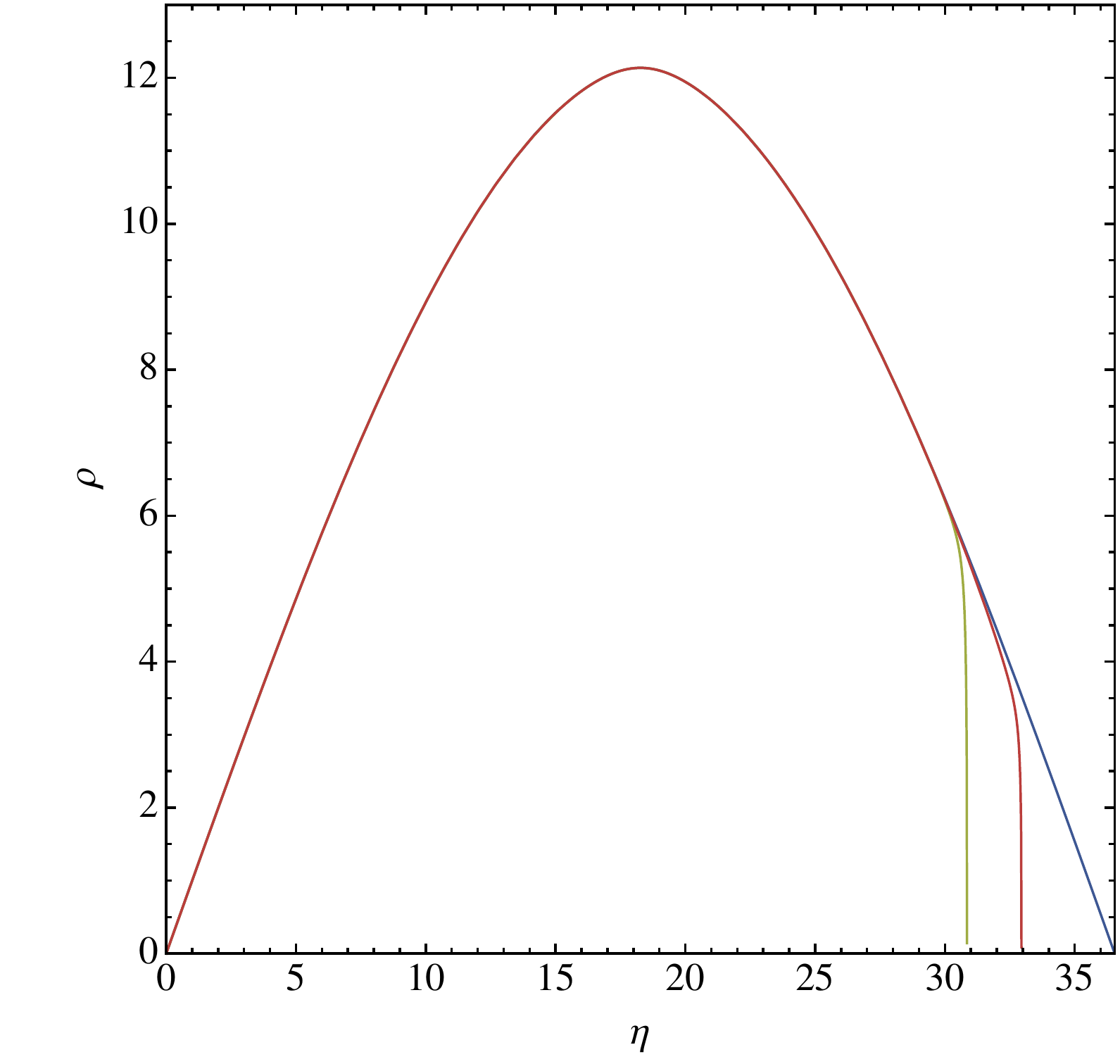}
\end{minipage}\vspace{0.4cm}\\
\begin{minipage}{\smallWidthLeft}
\includegraphics[width=\smallWidthRight]{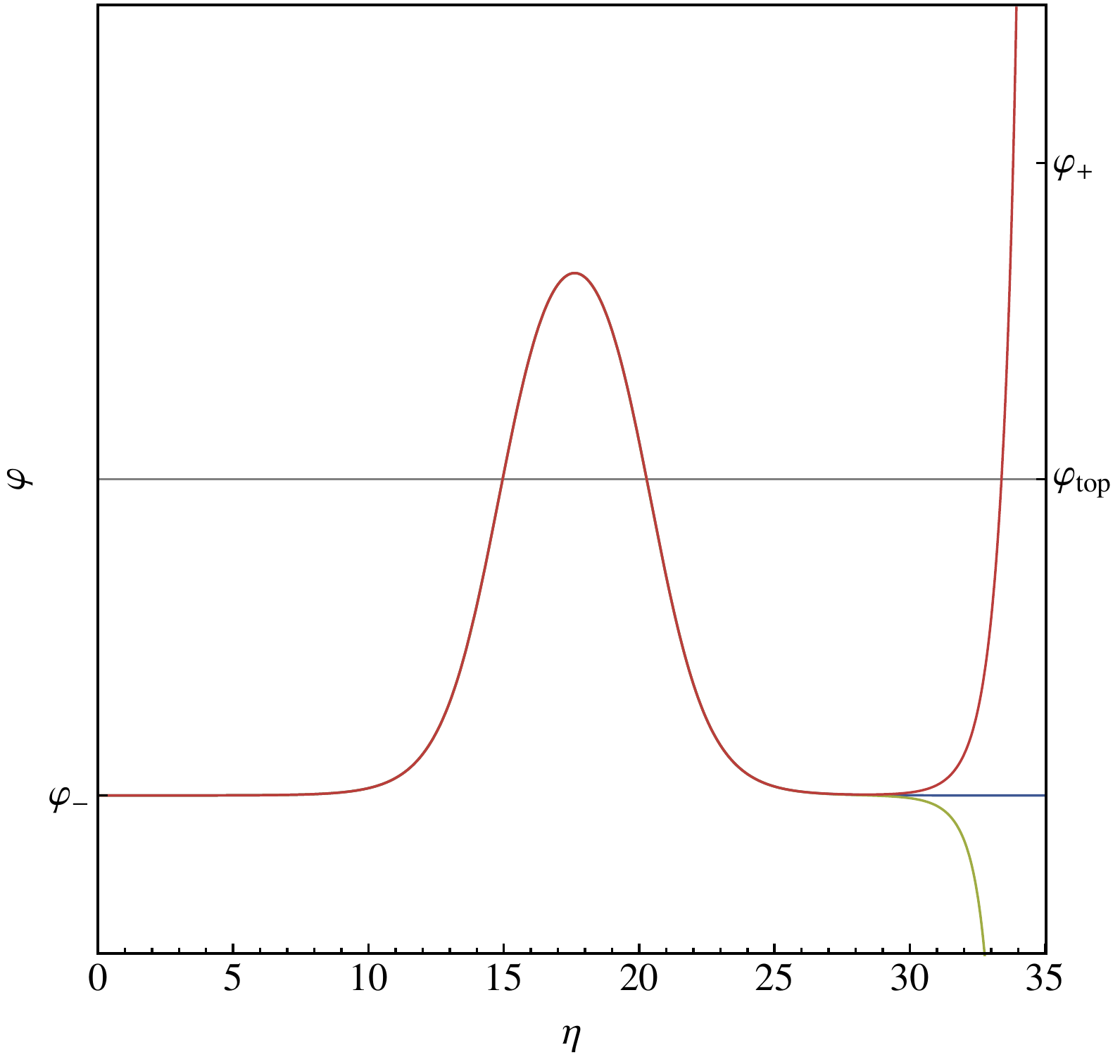}
\end{minipage}%
\begin{minipage}{\smallWidthRight}
\includegraphics[width=\smallWidthRight]{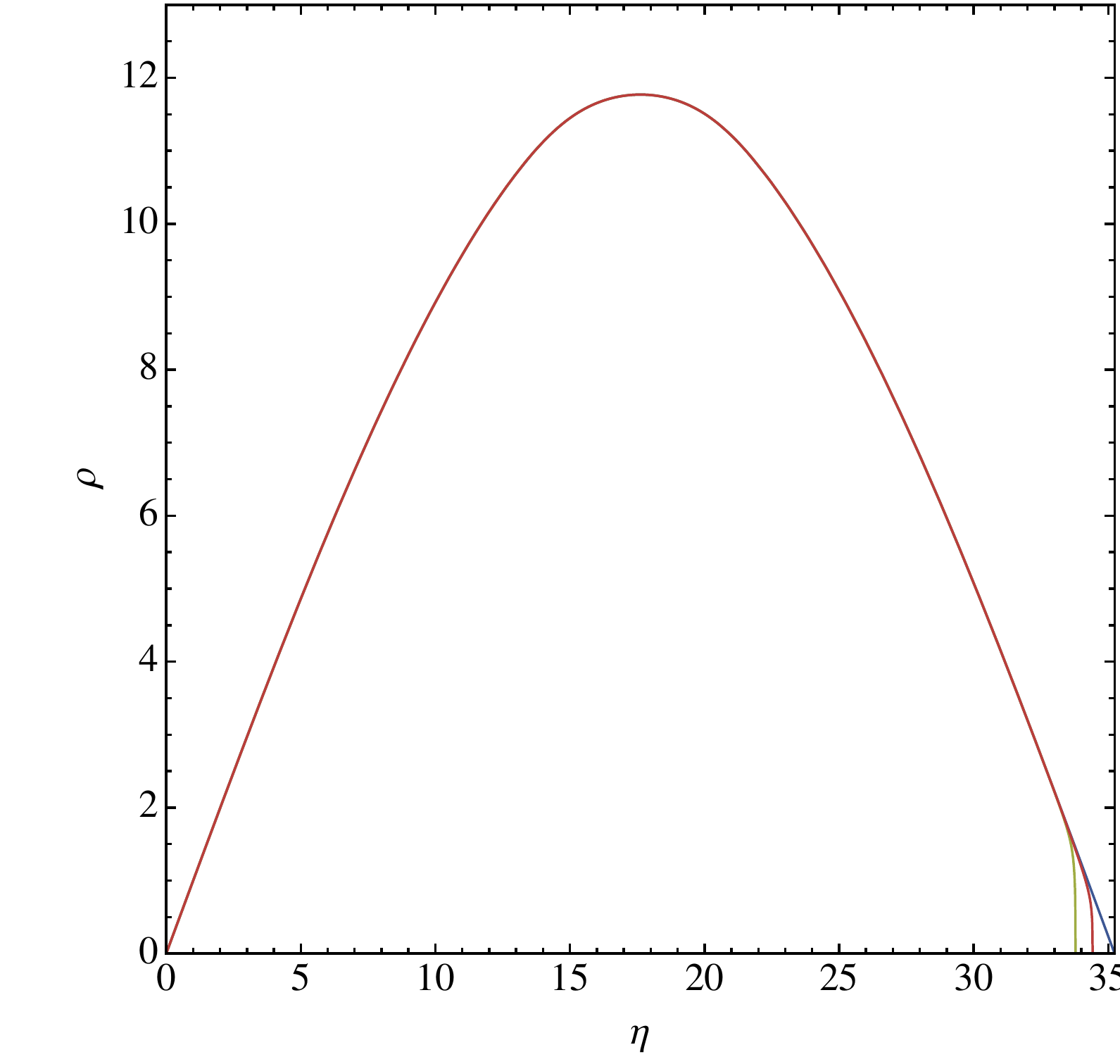}
\end{minipage}
\caption{\small  \label{fig:ladder} In the top panel, number of oscillations as a function of the starting
value of the scalar field $ \varphi_0$ ($V_0 = 0.5$, $ \kappa = 0.04$):
the values separating different classes of solutions correspond to regular instanton solutions.
In the center panels, profile of the scalar field and $ \rho$ for under/overshooting around the $N=1$ instanton.
In the bottom panels, same for $N=2$. As $ \varphi_0$ increases, the red profile of the center
panels evolves towards the green, then blue and red profiles of the bottom panels.}
\end{figure}\mbox{}

\subsection{Instantons in a Double Well Potential with Degenerate Minima}

In the limiting case $V_+ = V_-$ a Euclidean solution in flat spacetime does not exist. However, when gravity is included, instanton solutions between degenerate vacua become possible. Heuristically, this can be understood as follows: bounce solutions between dS vacua never start and end precisely at a minimum of the potential, but rather they start and end a little higher up on the potential. This can be interpreted by taking into account the fact that dS space has a non-zero temperature. Then, in a path integral formalism the preferred path connecting the original false vacuum to a bubble of true vacuum immersed in a region of false vacuum involves first a thermal excitation part of the way up the potential, followed by the bounce solution bringing the field over the hill. Because of this, the form of the bounce solution itself is insensitive to the behavior of the potential outside of the field range that it interpolates along. In other words, the precise height of the potential minimum is not directly relevant in determining the shape of the bounce solution, and tunneling ought to be able to take place irrespective of whether the minima are at slightly different or equal heights.

As an example of gravitational instanton solutions, we will consider the symmetric potential studied in the work of Lee {\it et al.} \cite{Lee:2011ms}, where such instanton solutions were presented:
\begin{equation} \label{eq:leePotential}
V( \varphi) = \frac{1}{8}( \varphi ^2 - 1) ^2 + V _0 \;.
\end{equation}
The two degenerate vacua are located at $ \varphi _{\pm} = \pm 1$, where $ V _{\pm} = V _0$. To reproduce the results
presented in \cite{Lee:2011ms}, we take $ V _0 = 0.5$, $ \kappa = 0.04$. From \eqref{eq:weinbergNumber} we find $ N = 6,$ suggesting that solutions with up to six interpolations of the scalar field should exist.
The results of the numerical solutions of the field equations are represented in Figure \ref{fig:ladder}.
The solutions that separate classes of profiles with the same number $N$ of oscillations are the regular bounce
solutions already described in \cite{Lee:2011ms}. In particular, the object separating singular solutions with $N$
oscillations and $N+1$ oscillations is a regular bounce with $N$ oscillations. Such regular solutions appear to be
always symmetric ($N$ even) or antisymmetric ($N$ odd) under $ \eta \rightarrow \bar{ \eta} - \eta$. This result looks
natural, considering the symmetry of the potential, but is not predicted by any mathematical proof that we are aware of.

We would like to stress that the existence of these solutions is not immediately apparent from the original formalism developed by Coleman and De Luccia. Much of their work assumes that the energy difference between the two minima is small, yet non--zero, in such a way that the bounce solution can be effectively thought of as a bubble of true vacuum separated by a thin wall from the false vacuum. In the case of compact solutions, a new scale enters the picture, namely the Hubble scale, and one must be careful in defining the thin wall limit.

\begin{figure}[t]
\begin{minipage}{\smallWidthLeft}
\includegraphics[width=\smallWidthRight]{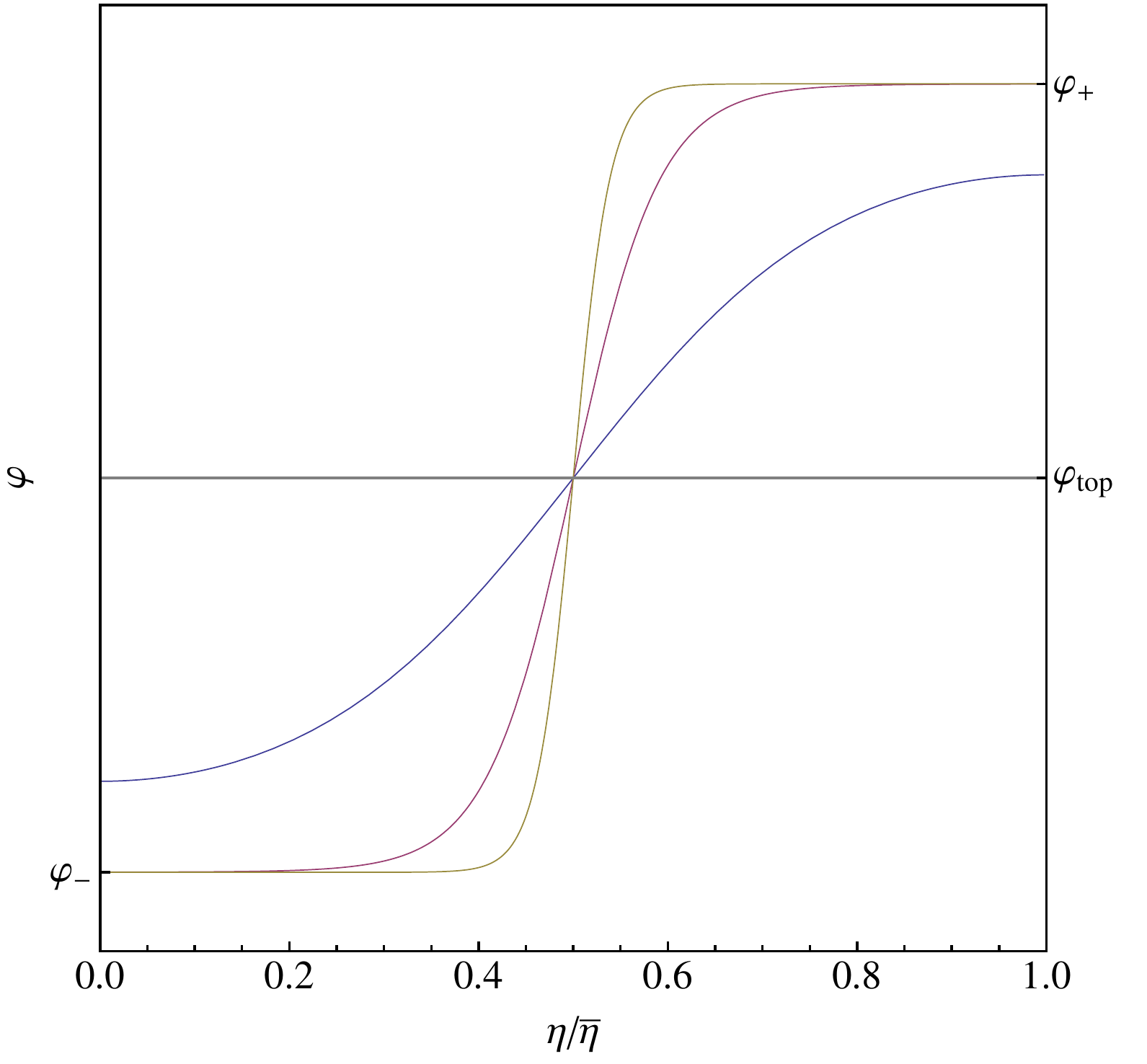}
\end{minipage}%
\begin{minipage}{\smallWidthRight}
\includegraphics[width=\smallWidthRight]{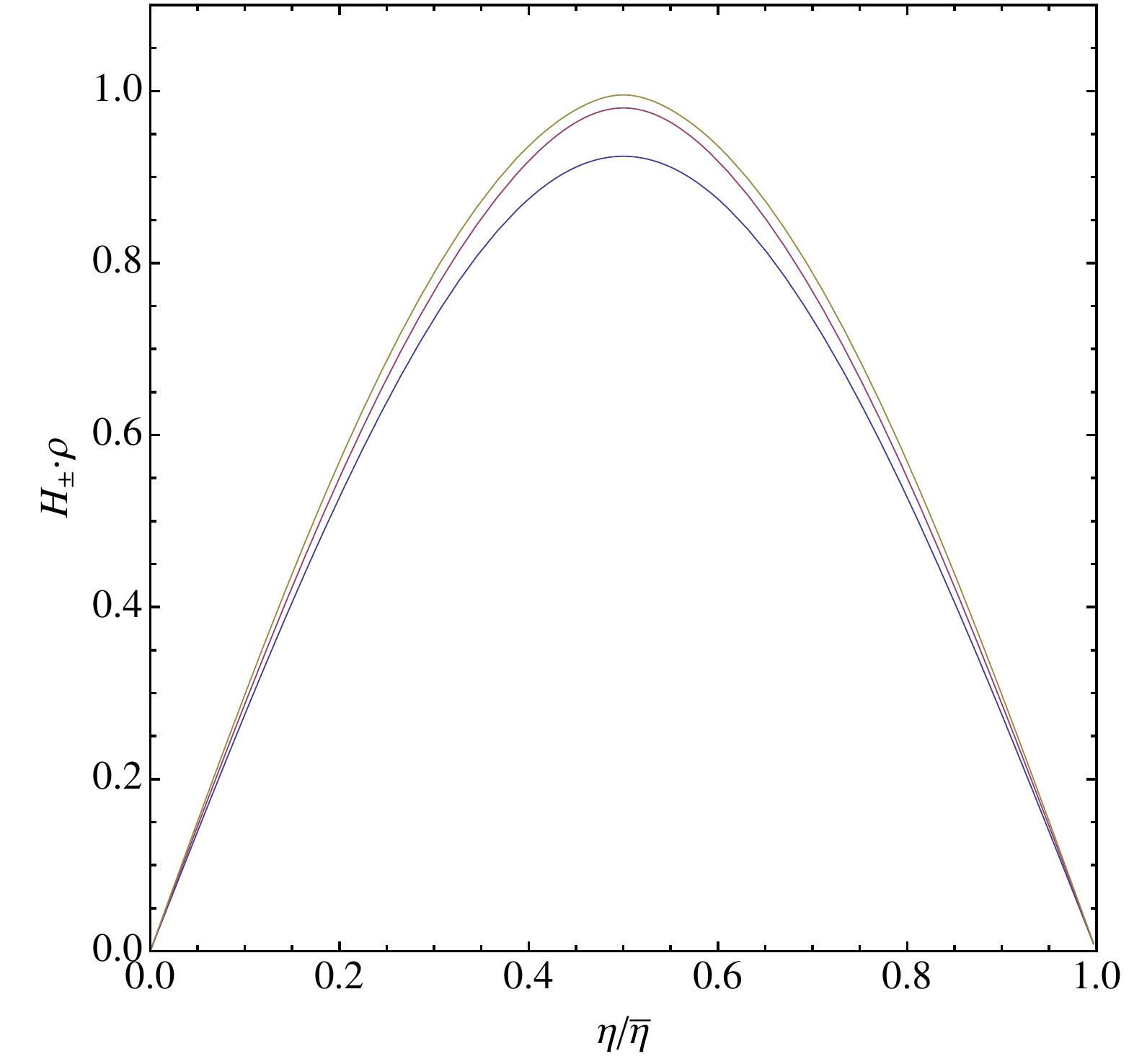}
\end{minipage}\caption{\small \label{fig:thinWall} The left panel shows the profiles of the first instanton
for the potential in \eqref{eq:leePotential} with $V_0 = 0.5$,
 for the values $ \kappa = 0.4$ (blue line),
$ \kappa = 0.09$ (purple line) and $ \kappa = 0.02$ (beige line). In the right panel, the corresponding profiles of the
$ \rho$ function are plotted. As $ \kappa \rightarrow 0$, the wall position scales as the Hubble radius $H _0 ^{-1} \propto \kappa ^{-1/2}$, while its thickness
remains approximately constant.}
\end{figure}

In flat space the thin wall limit is defined by the condition
\begin{equation} \label{eq:thinWallOld}
S/(V _{+} - V _{-}) \gg  |V_{, \varphi \varphi}( \varphi_{top})|^{-1/2} \;,
\end{equation}
where $S$ is a constant, of dimensions  $ (\textrm{length})^{-3}$, which characterizes the potential barrier. When the difference in energy density between the two vacua goes to zero, the size of the bubble diverges while the
thickness of its wall remains approximately constant. In de Sitter space the situation is radically different:
as we recalled above, bounce solutions survive when the two vacua become degenerate. If we require the wall to be thin
compared to the bubble size we need, in particular, its thickness to be small compared to the size of the whole bounce
geometry, which is set by $H _0 ^{-1}$. With this assumption, it is easy to deduce from the overshooting--undershooting argument
that the wall should be located close to the maximum of $ \rho( \eta)$, where the coefficient $ \rho'/ \rho$ vanishes.
Therefore, in the case of degenerate vacua, the thin--wall regime is attained when the following inequality holds:
\begin{equation} \label{eq:thinWallNew}
\eta_{wall} \sim H _{ 0} ^{-1}  \gg |V_{, \varphi \varphi}( \varphi_{top})|^{-1/2} \;.
\end{equation}
In other words, the radius of the bubble approaches the Hubble radius as the energy difference between the two vacua
tends to zero (see Figure \ref{fig:thinWall}).

Also, in the standard treatment of Coleman and De Luccia, the existence of a bubble wall is physically justified by the fact that it carries the difference in energy between the true and false vacua. In a sense, the overall energy is conserved in the tunneling process, and the energy that is lost by forming a bubble of true vacuum is compensated by the energy of the wall. In our case, the interior and the exterior of the bubble consist of vacua which have the same energy density, and so one may wonder how
a wall, carrying tensile energy, is possible at all. The resolution is that, in the limit that the vacua become degenerate, a new effect becomes apparent\footnote{Presumably this effect is also present, though likely increasingly subdominant, when the vacua are taken to be at heights that differ more and more.}. In fact, what we find is that the presence of the wall is compensated by the removal of a section of the Euclidean four sphere. Let us elaborate on this statement: the wall is approximately located at the equator of the four sphere, where
\begin{equation}
\rho \simeq H _0 ^{-1} \;.
\end{equation}
Let $ \eta_A < \eta_B$ such that the wall is completely included in the region $ \eta_A \leq \eta \leq \eta_B$ and
effectively $V \simeq V_0$ outside of this interval. We then have
\begin{eqnarray}
\rho & \simeq & H _0 ^{-1}\, \sin{ \left( H _0 \eta \right)}: \quad \eta \leq \eta_A \;,\\ \label{eq:coincidence}
\rho & \simeq & H _0 ^{-1}\, \sin{ \left(H _0 (\eta + \Delta \eta) \right)}: \quad \eta \geq \eta_B \;.
\end{eqnarray}
\begin{figure}[t]
\begin{minipage}{\smallWidthLeft}
\includegraphics[width=\smallWidthRight]{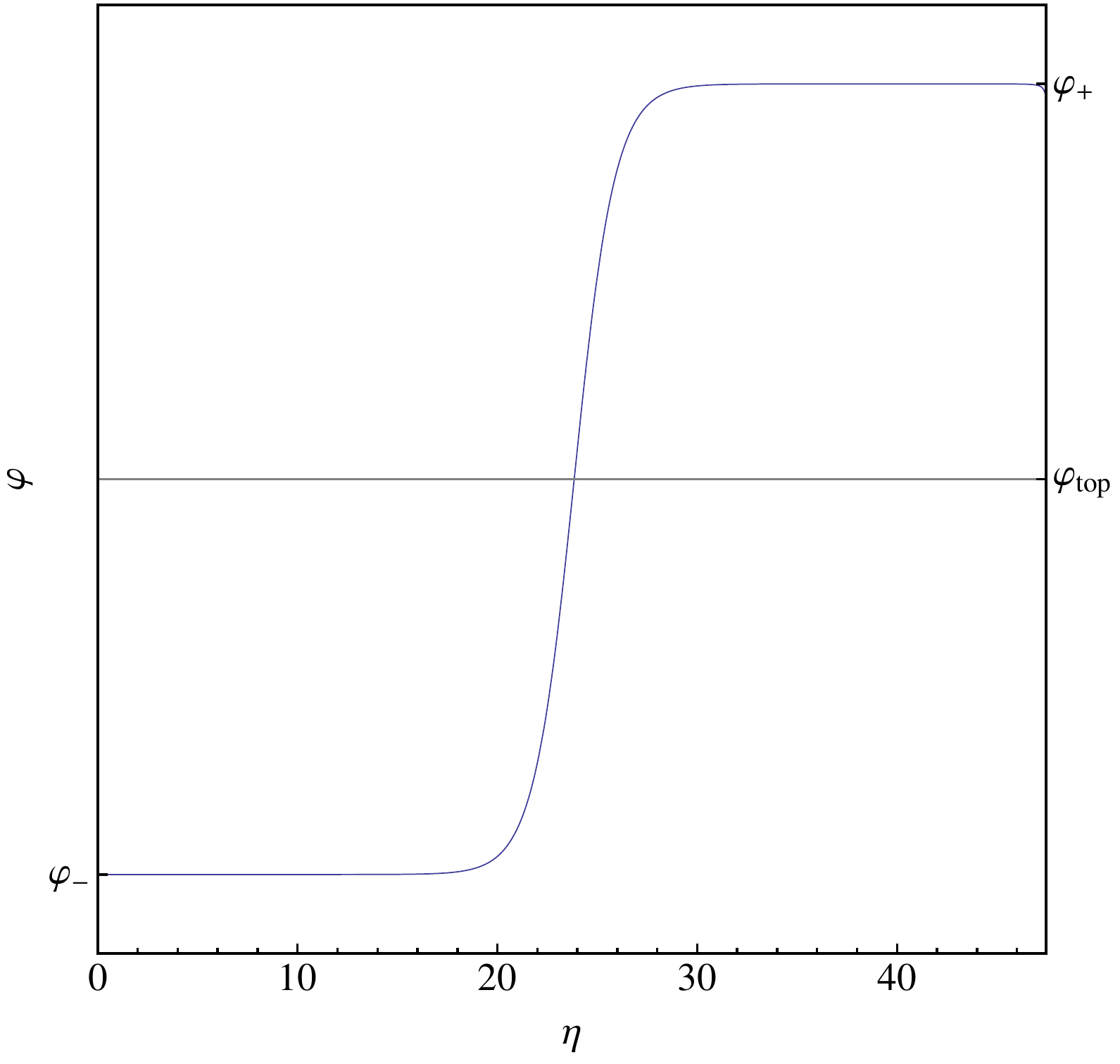}
\end{minipage}%
\begin{minipage}{\smallWidthRight}
\includegraphics[width=\smallWidthRight]{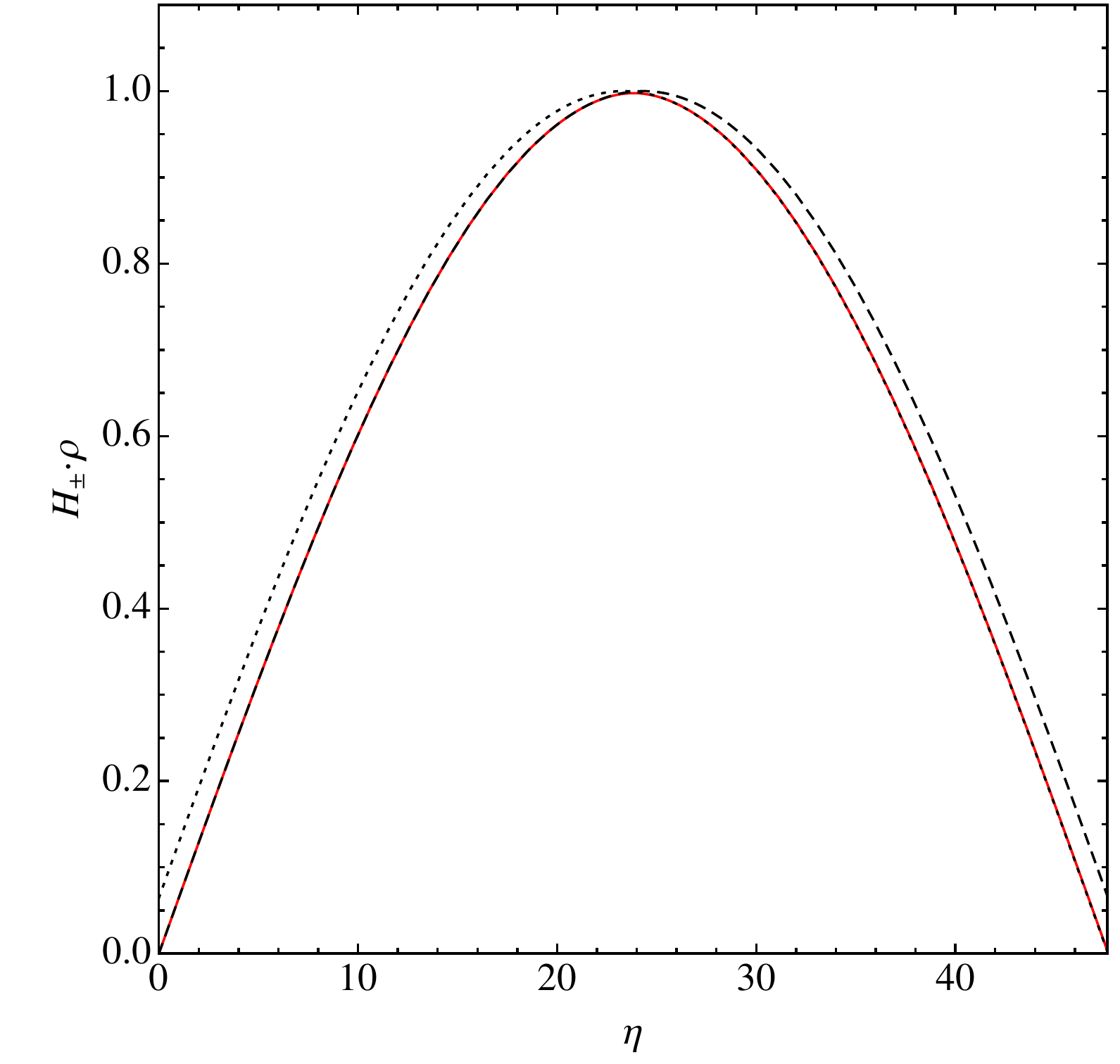}
\end{minipage}%
\caption{\small \label{fig:aBitAway} Left panel: the scalar field profile for the first, $N=1$, instanton solution
with $V_0 = 1$ and $ \kappa = 0.0125$. Right panel: the red, solid line represents the corresponding
profile of $ \rho( \eta)$. The black, dashed line represents the function $ \rho_0$ describing the vacuum
de Sitter solution $ \varphi = \pm 1$. The black, dotted line corresponds to the shifted de Sitter
solution \eqref{eq:coincidence} with $ \Delta \eta$ estimated according to \eqref{eq:deltaEta}.}
\end{figure}%
As we argued above, the thin wall should be located at the equator of the four sphere, where $ \rho \simeq H _0 ^{-1}$.
If we take $ \eta_B - \eta_A$ much smaller than $H _0^{-1}$, from the field equations we get
\begin{equation}
\rho ^{ \prime \prime} \simeq - \frac{ \kappa H _0 ^{-1}}{3} \left( \varphi ^{ \prime 2} + V \right),
\quad \eta \in [ \eta _{A}, \eta _{B}] \;.
\end{equation}
We can now estimate the difference
\begin{equation}
\rho'( \eta_B) - \rho_0'( \eta_B) \simeq - \frac{ \kappa H _0 ^{-1}}{3}
\int_{ \eta_A}^{ \eta_B} d \eta \left( \varphi ^{ \prime 2} + V - V_0 \right) \;.
\end{equation}
As in the standard thin wall limit, the friction/anti--friction term in the scalar field equation can usually be neglected at the wall:
\begin{equation}
\frac{1}{2} \varphi ^{ \prime 2} - V \simeq -V _0 \;.
\end{equation}
Comparing with \eqref{eq:coincidence} we find that
\begin{eqnarray}
H _0 \Delta \eta & \simeq & \frac{ \kappa H _0 ^{-1}}{3}
\int_{ \eta_A}^{ \eta_B} d \eta \left( \varphi ^{ \prime 2} + V - V_0 \right) \\
& \simeq & \kappa H _0 ^{-1} \int_{ \varphi_{-}} ^{ \varphi ^{+}} d \varphi \, \sqrt{ \frac{V - V _0}{2}} \;.
\end{eqnarray}
that is
\begin{equation}  \label{eq:deltaEta}
\Delta \eta \simeq \frac{3}{ \sqrt{2}V _0} \int_{ \varphi_{-}} ^{ \varphi ^{+}} d \varphi \, \sqrt{ V - V _0} \;.
\end{equation}
This formula gives the width of the slice of four--sphere that is removed due to the presence
of the wall. Numerical results for the potential \eqref{eq:leePotential} indicate a good agreement
with this analytic consideration (see Figure \ref{fig:aBitAway} for an illustration). Thus, loosely speaking, the wall is created by the conversion of a slice of spacetime into scalar field gradient energy.

\section{Spectrum of Linear Perturbations}

As discussed above, the study of the spectrum of perturbations about a given Euclidean solution is important
in order to determine whether the solution actually contributes to the decay rate of a given vacuum.
Indeed, the presence of a single perturbation mode with a negative eigenvalue is required for this
interpretation \cite{Coleman:1987rm}.
The negative mode of the Coleman - De Luccia bounce was first found in
\cite{Khvedelidze:2000cp,Lavrelashvili:1999sr} and the fluctuation spectrum about oscillating bounces was studied in
\cite{Lavrelashvili:2006cv,Dunne:2006bt}.
In the case of bounce solutions (connecting two {\it non--degenerate} dS vacua),
numerical evidence showed that the $N^{\rm th}$ oscillating bounce is characterized
by $N$  homogeneous, $O(4)$--symmetric, negative modes. For $N >1$, there also exist additional inhomogeneous
negative modes with non-zero angular quantum number $ \ell > 0$.

In their study of oscillating instanton solutions connecting degenerate vacua,
the authors of \cite{Lee:2009bp,Lee:2011ms}, probably based on the properties of instantons in quantum mechanics,
argued that such solutions should not admit negative modes.
In what follows we address this question numerically  and
show that this expectation is not fulfilled.

The quadratic action for linear $O(4)$--symmetric perturbations about Euclidean solutions
in a self-interacting scalar field theory coupled to gravity
was derived in \cite{Khvedelidze:2000cp,Lavrelashvili:2006cv}, and reads:
\be
S^{(2)}_E= 2 \pi^2 \int \left( \frac{1}{2} {f'}^2 + \frac{1}{2} U[\rho(\eta), \varphi(\eta)] f^2 \right) d\eta \kma
\ee
with the potential
\begin{eqnarray}
U[\rho(\eta), \varphi(\eta)] & \equiv &  \frac{1}{ \mathcal{ Q}} V_{, \varphi \varphi} - \frac{10 \rho ^{ \prime 2}}{ \rho ^2 \mathcal{Q}}
+ \frac{ 12 \rho ^{ \prime 2}}{ \rho ^2 \mathcal{ Q} ^2} + \frac{8}{ \rho ^2 \mathcal{Q}} - \frac{ 6}{ \rho ^2}
- \frac{ 3 \mathcal{Q}}{ \rho ^2} - \frac{ \rho ^{ \prime 2}}{4 \rho ^2} \nonumber \\
&&+ \frac{ \kappa \rho ^2}{2 \mathcal{Q} ^2} V _{, \varphi}^2 - \frac{ 2 \kappa \rho \rho' \varphi'}
{ \mathcal{Q} ^2} V_{, \varphi} - \frac{ \kappa}{6} \left( \varphi ^{ \prime 2} + V \right) \kma
\end{eqnarray}
where $f \equiv \frac{ \rho ^{3/2}}{ \mathcal{Q}} \delta \varphi$ represents the fluctuation of a scalar field and
$\mathcal{Q}  \equiv  1- \frac{k \rho ^2 \varphi ^{ \prime 2}}{6}$.
The Schr\"odinger equation diagonalizing the above quadratic actions reads
\be \label{eq:schrodinger}
- f ^{ \prime \prime} + U[\rho(\eta), \varphi(\eta)] f  =  E f \pkt
\ee
Now our aim is to study the number of negative modes in the spectrum of \eqref{eq:schrodinger}
for different oscillating instantons.
A simple method of counting states with negative eigenvalues of the Schr\"odinger equation
consists in investigating the zero energy wavefunction of this equation.
According to well known theorems \cite{aq95}, the number of nodes of the zero energy wavefunction then counts
the number of bound states in a given potential (heuristically, this is because the number of nodes points to the existence of a corresponding number of lower energy wavefunctions with successively fewer nodes).

Solving numerically the background equations for the oscillating bounce solution
in the potential \eqref{eq:leePotential} and the perturbation equations with appropriate initial conditions
(see \cite{Lavrelashvili:2006cv} for details)
for wide class of investigated solutions we find that the instanton with $N$ nodes
has exactly $N$ homogeneous ($O(4)$--symmetric) negative modes.

\begin{figure}[thbp]
\begin{minipage}{\thirdWidthLeft}
\includegraphics[width=\thirdWidthRight]{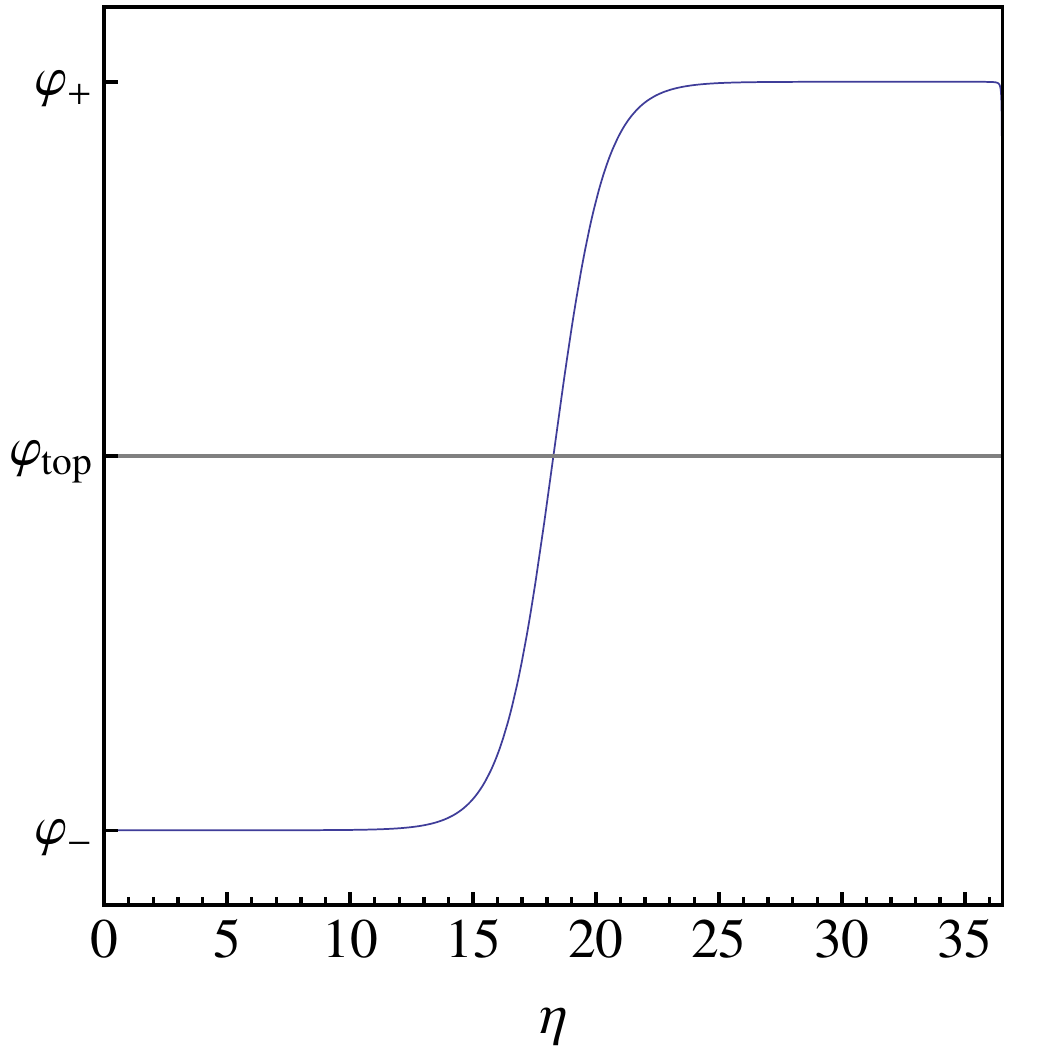}
\end{minipage}%
\begin{minipage}{\thirdWidthLeft}
\includegraphics[width=\thirdWidthRight]{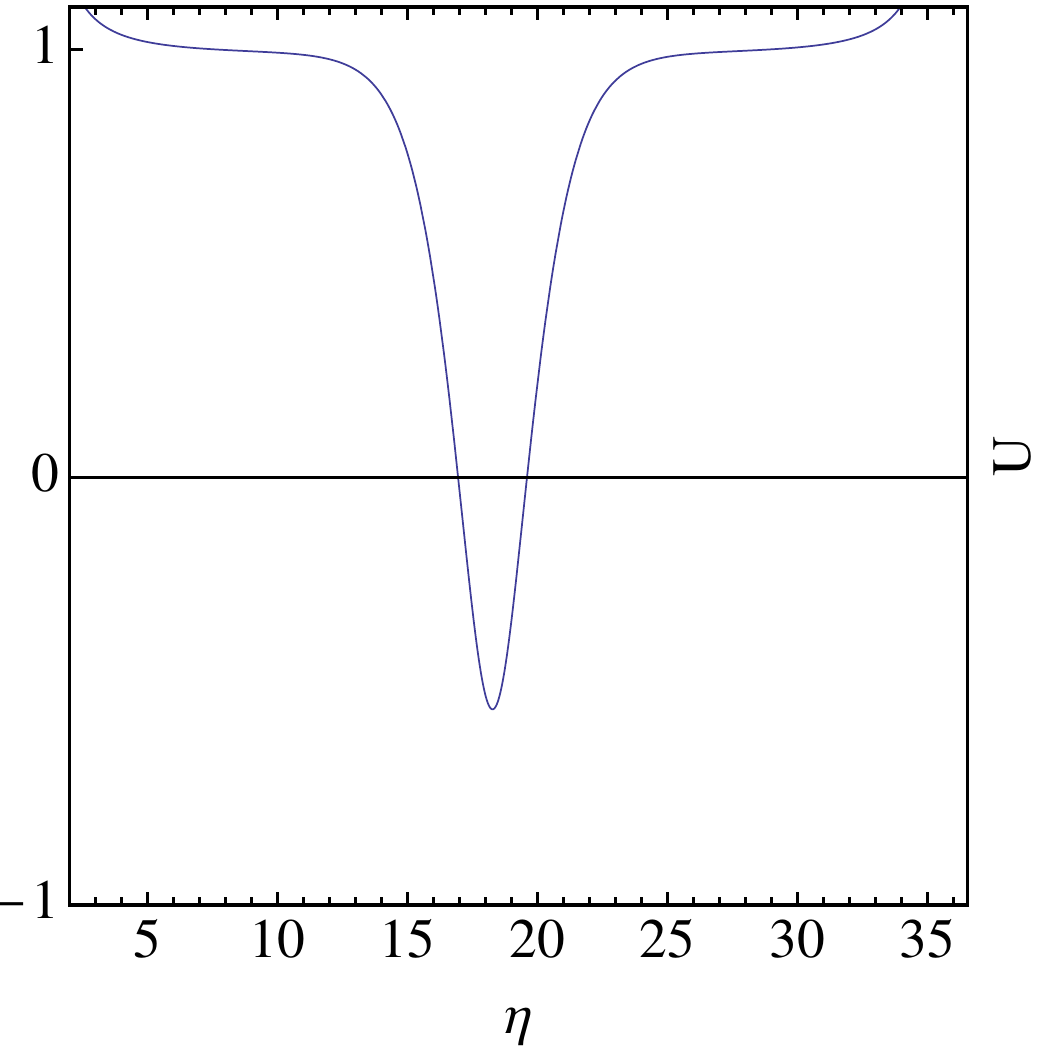}
\end{minipage}%
\begin{minipage}{\thirdWidthRight}
\includegraphics[width=\thirdWidthRight]{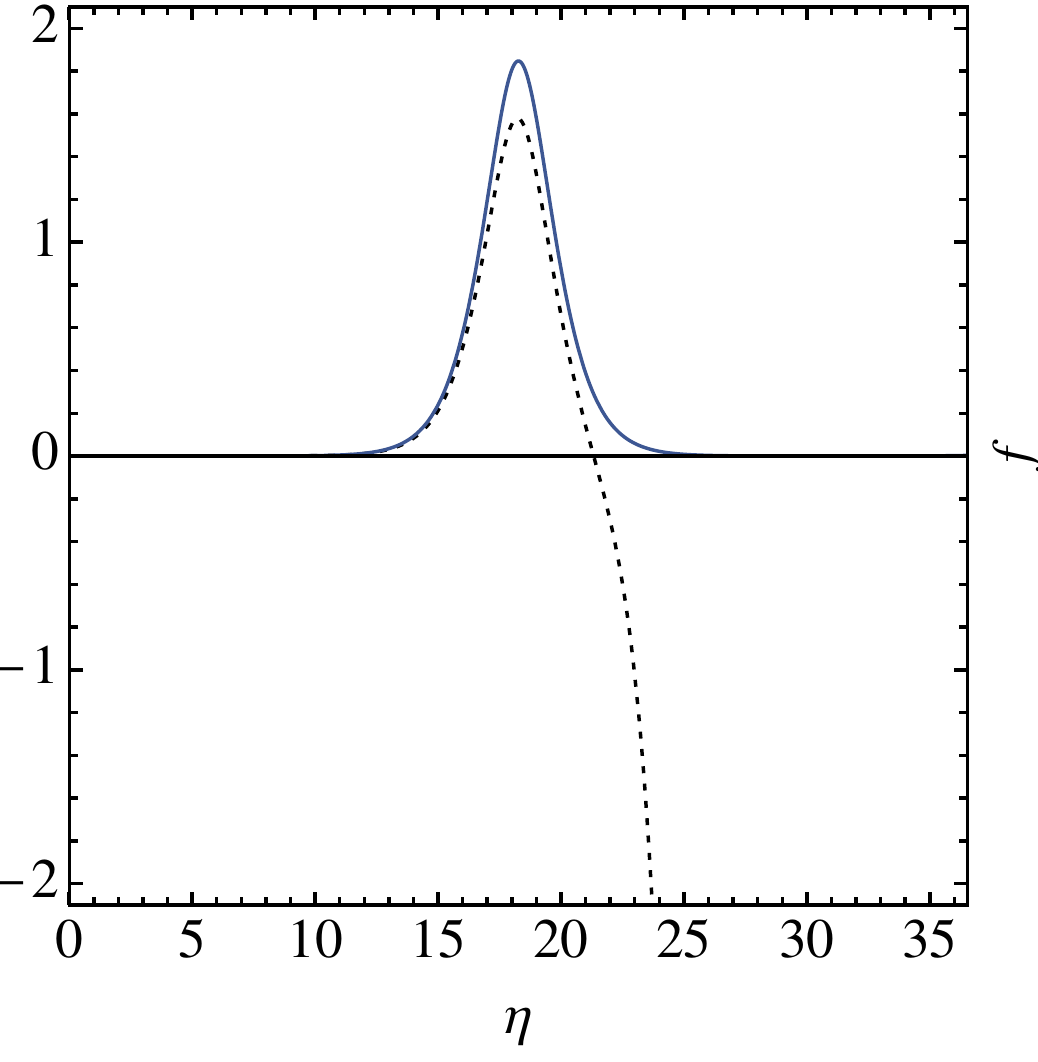}
\end{minipage}
\vspace{0.3cm}\\
\begin{minipage}{\thirdWidthLeft}
\includegraphics[width=\thirdWidthRight]{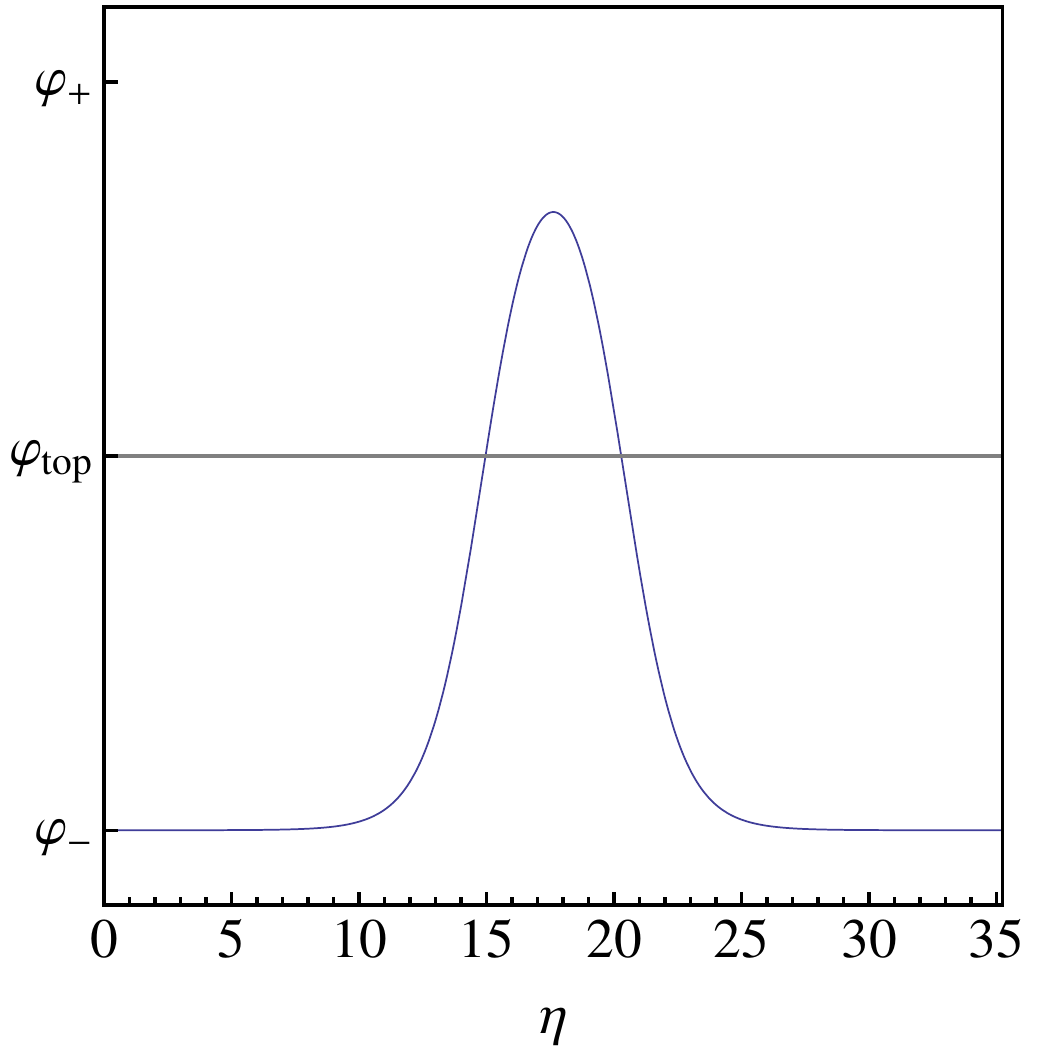}
\end{minipage}%
\begin{minipage}{\thirdWidthLeft}
\includegraphics[width=\thirdWidthRight]{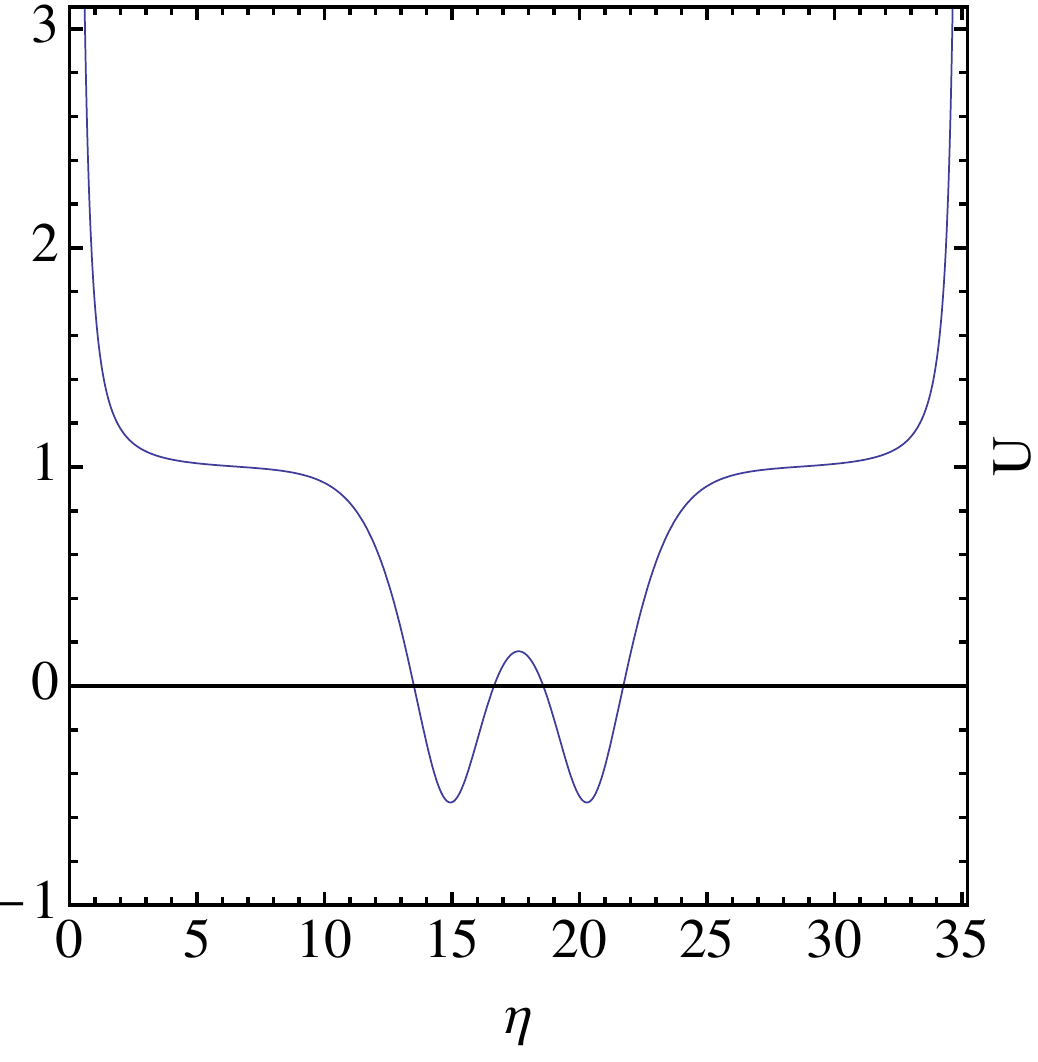}
\end{minipage}%
\begin{minipage}{\thirdWidthRight}
\includegraphics[width=\thirdWidthRight]{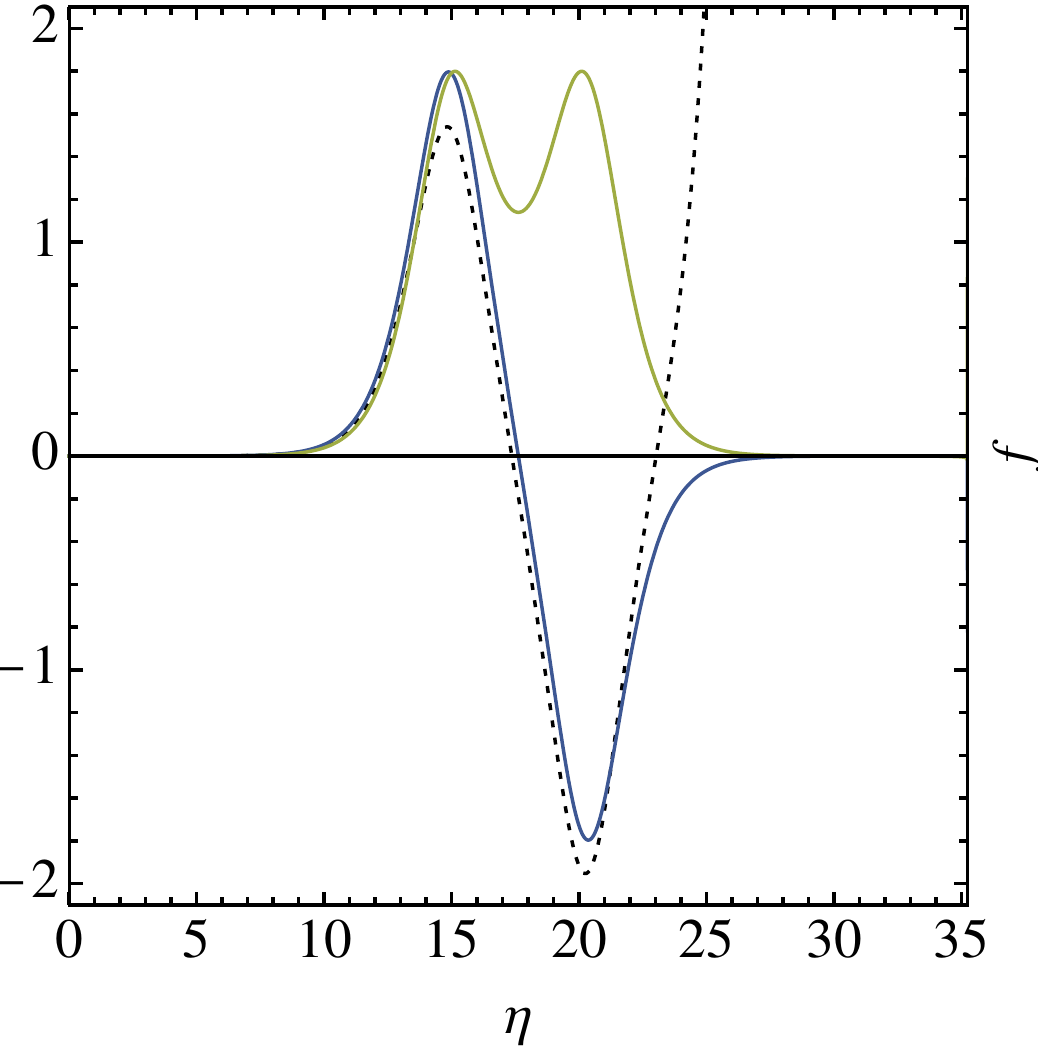}
\end{minipage}\vspace{0.3cm}\\
\begin{minipage}{\thirdWidthLeft}
\includegraphics[width=\thirdWidthRight]{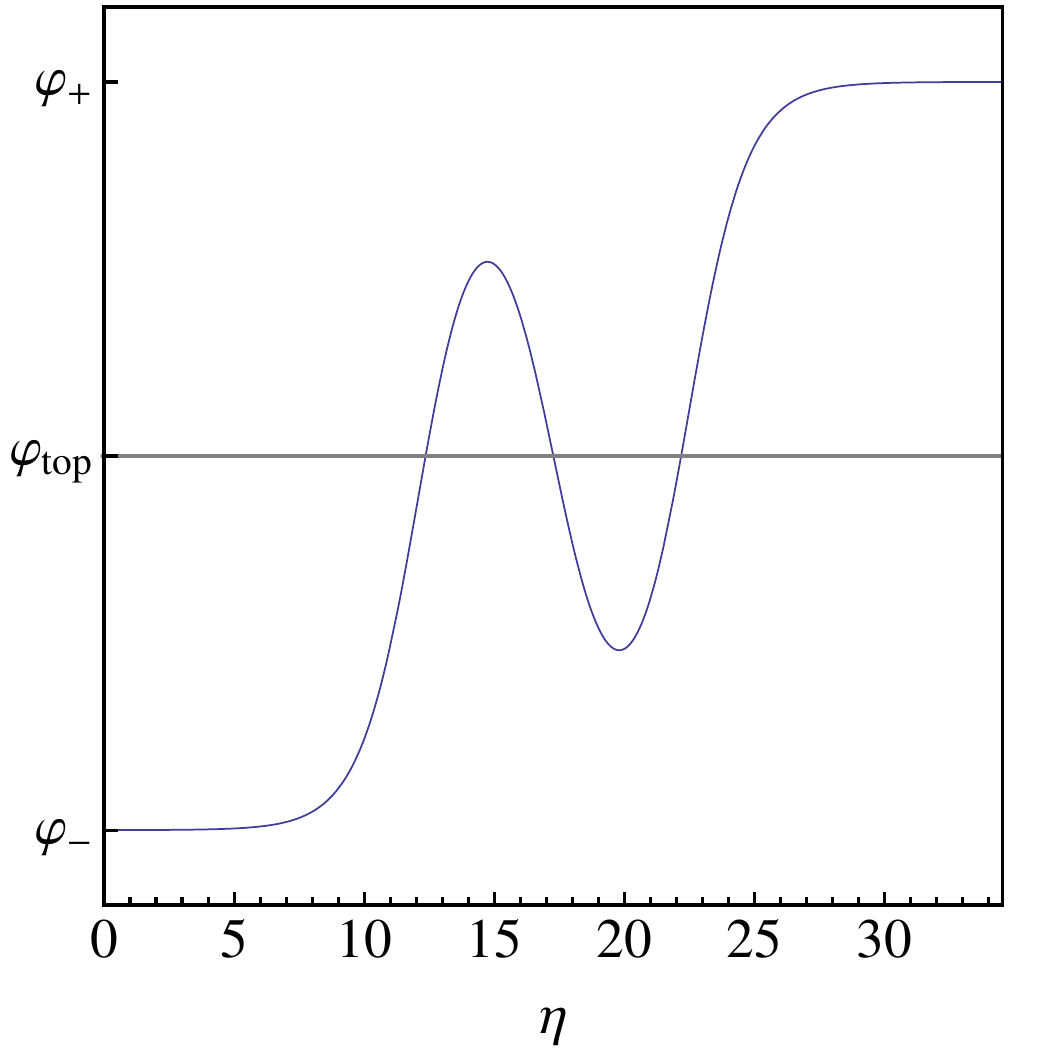}
\end{minipage}%
\begin{minipage}{\thirdWidthLeft}
\includegraphics[width=\thirdWidthRight]{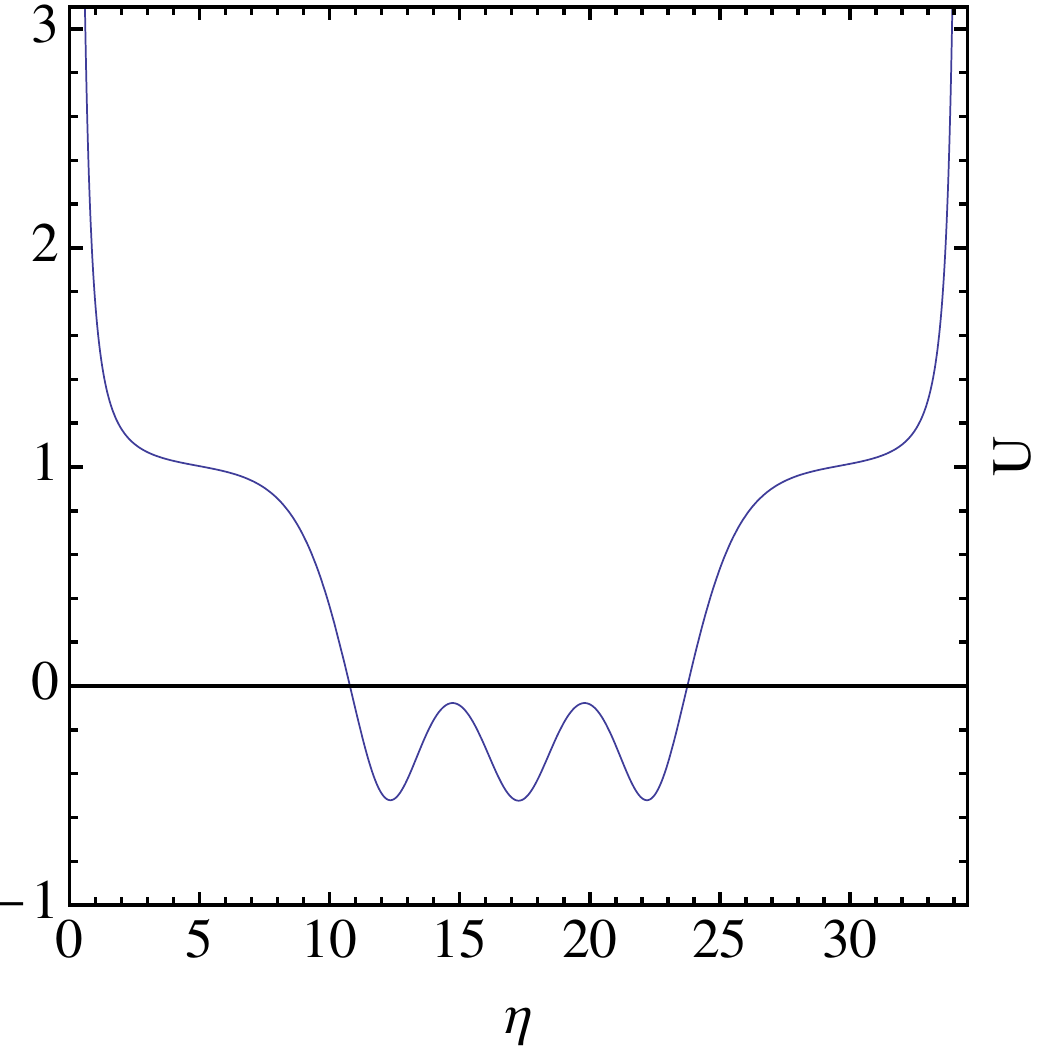}
\end{minipage}%
\begin{minipage}{\thirdWidthRight}
\includegraphics[width=\thirdWidthRight]{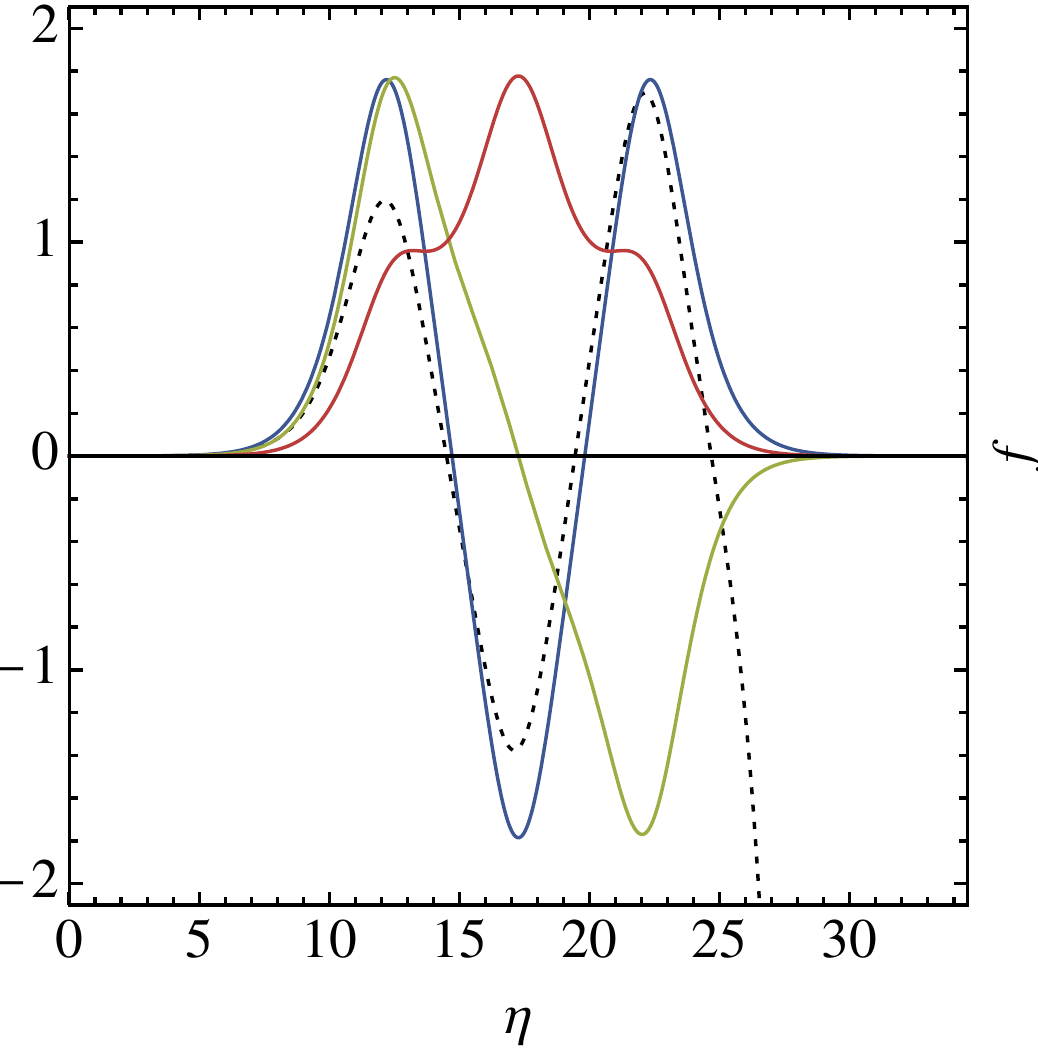}
\end{minipage}
\caption{ \small \label{fig:modesdSOne}
In the left panels, profile of the scalar field for the first three oscillating instantons with $N=1,2,3$ and
$ V_0 = 0.5$ and $ \kappa = 0.04$. In the central panels, potential for $O(4)$--symmetric perturbations. In the right panels, zero mode wavefunction (dotted line) and negative modes (solid lines): the normalization of the wavefunctions is not imposed, so the overall scale of the vertical axis is irrelevant. For $N=4,5,6$ we found analogous results.}
\end{figure}

First, we considered the bounce solutions connecting two degenerate dS vacua.
Our findings are illustrated in Figure \ref{fig:modesdSOne}.
For definiteness, we have taken the same parameter values as those considered in \cite{Lee:2011ms},
namely $V_0 = 0.5$ and $ \kappa = 0.04$, for which six oscillating bounce solutions are possible.
The profile of the potential $U( \eta)$ for the first three solutions is represented in Figure \ref{fig:modesdSOne}.
It is interesting to note that around the $N$--oscillating solution, the potential $U$ admits $N$ negative minima.
The corresponding zero energy wavefunction shows exactly $N$ nodes\footnote{Generally speaking, the precise shape of modes and the corresponding eigenvalues depend on the choice
of the weight function used to define the eigenvalue problem.
However, the number of negative energy eigenstates is independent of this choice.
Therefore, adopting {\it e.g.} the weight function of \cite{Dunne:2006bt},
one obtains again the same number of negative modes,
though in general they differ from the ones computed according to \eqref{eq:schrodinger}.}
(right panels of Figure \ref{fig:modesdSOne}).

\begin{figure}[h]
\begin{minipage}{\smallWidthLeft}
\includegraphics[width=\smallWidthRight]{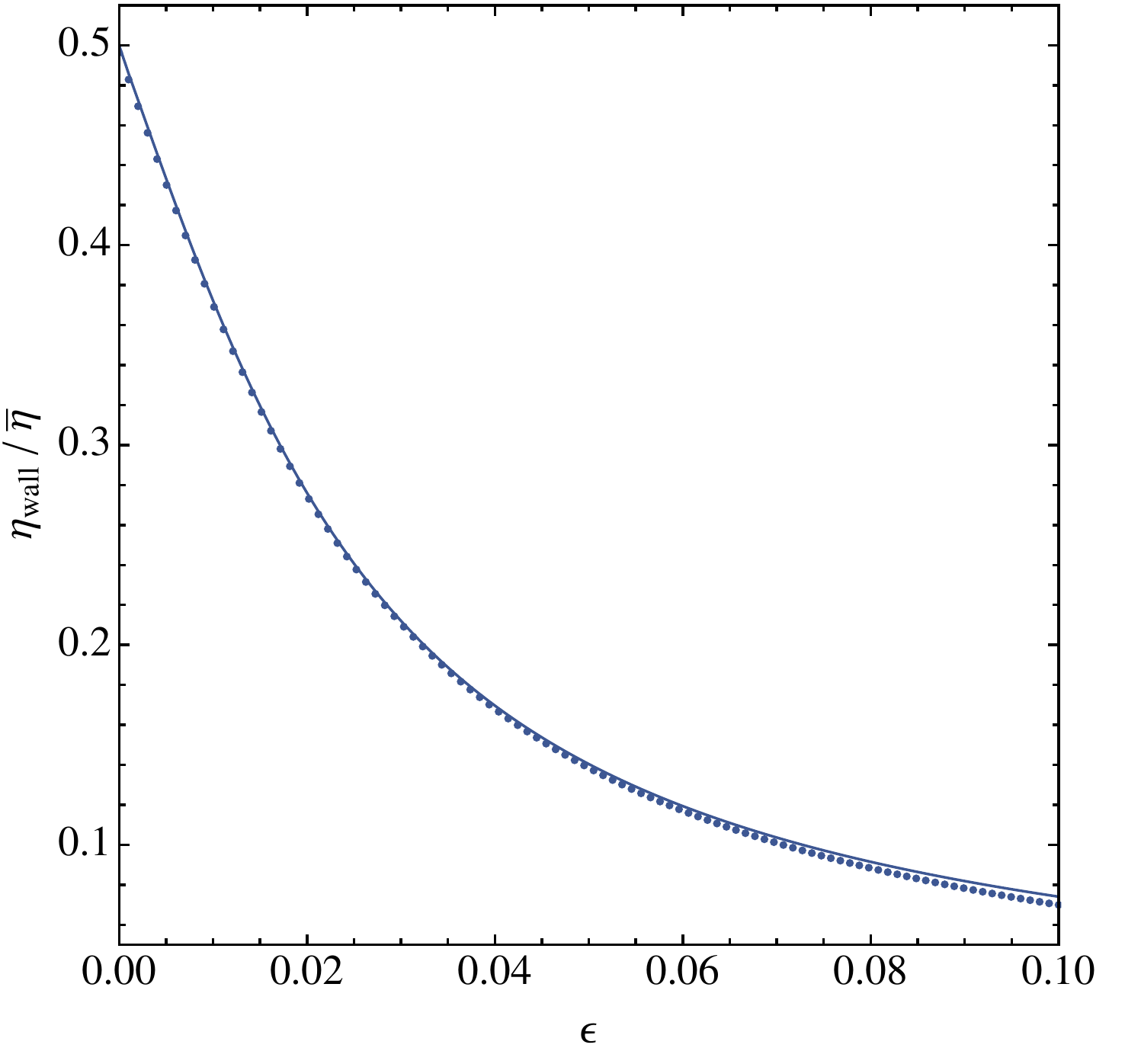}
\end{minipage}%
\begin{minipage}{\smallWidthRight}
\includegraphics[width=\smallWidthRight]{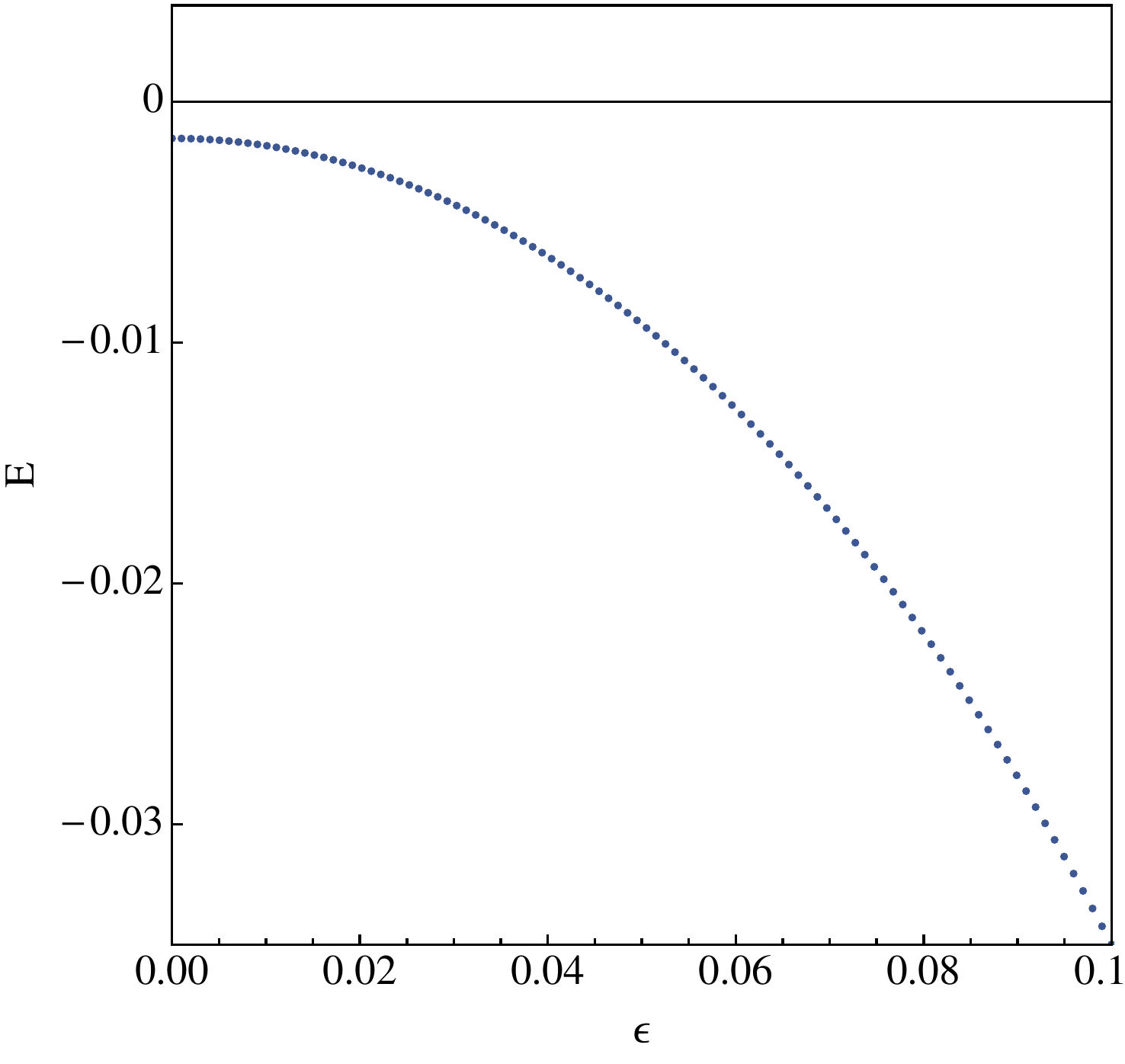}
\end{minipage}%
\caption{\small \label{fig:degenerateLimit} Left panel: position of the wall of the $N=1$ bounce normalized to the size $ \bar{ \eta}$ of the de Sitter sphere, as a function of the energy difference between the two vacua, for $V_0 = 0.5$ and $ \kappa = 0.4$ (dots). $ \eta_{wall}$ was conventionally defined requiring $ \varphi( \eta_{wall}) = \varphi_{top}$. The solid line is the wall position as predicted by the Coleman--de Luccia treatment. When the two vacua are degenerate, the wall is located at the equator of the sphere ($ \eta_{wall}/ \bar{ \eta} = 1/2$). Right panel: eigenvalue of the $O(4)$ symmetric negative mode associated to the $N=1$ bounce.}
\end{figure}

Our findings suggest that the results obtained in \cite{Lavrelashvili:2006cv,Dunne:2006bt} do not
undergo significant changes when the energy difference between the two vacua vanishes.
This can be made explicit by numerically studying the degenerate limit. To do this, we add a linear term to the potential \eqref{eq:leePotential}:
\begin{eqnarray}
 \label{eq:leePotentialModified}
V( \varphi) & = & \frac{1}{8}( \varphi ^2 - 1) ^2 + V _0 + \epsilon \varphi \;, \\
V _{+} - V _{-} & = & 2 \epsilon + \mathcal{O}( \epsilon ^2) \;,
\end{eqnarray}
and investigate the limit $ \epsilon \rightarrow 0$. The results of this analysis for a particular value of $ V_0$ and $ \kappa$ are summarized in Figure \ref{fig:degenerateLimit}. We chose the parameters in such a way that the thin wall approximation applies to the bounce/instanton solutions. When $ \epsilon \rightarrow 0$, the wall moves towards the equator of the $4$--sphere. However, in the same limit, the negative eigenvalue remains finite, and reaches the instanton value. Finally, the wall position can be predicted by the Coleman--de Luccia treatment, keeping in mind that the vacuum energy is now finite.

We also extended our study  to bounce solutions connecting Minkowski and AdS degenerate vacua,
which  were also discussed in \cite{Lee:2011ms}. As it stands, our formalism only applies to the cases where the function $Q$ is positive. The cases that we investigated are similar in spirit to the instanton solutions connecting dS vacua, in that the instantons themselves are compact due to the positive value of the potential at $ \varphi_{top}$. It is then not surprising that similar results are obtained for these cases, and again $N$-oscillating solutions possess $N$ negative modes. It would be interesting to see if one can extend this class of solutions to the case where they start and end at zero or negative values of the potential. We leave this question for future work.

\section{Conclusions}

In the present paper we were interested in the properties of oscillating
instanton solutions in a scalar--gravity theory.
We investigated linear perturbations about instantons and studied numerically the corresponding Schr\"odinger equation.
Our results imply that instanton solutions with $N$ nodes admit exactly $N$ homogeneous negative modes.
The existence of additional negative modes for the oscillating instantons with more than one node
gives us ground to discard them as physical final configurations. Following the interpretation of \cite{Brown:2007sd}, they should rather be interpreted as
unstable intermediate thermal configurations interpolating between the basic instanton and Hawking-Moss solutions,
similarly to the oscillating bounces \cite{Brown:2007sd}. In contrast to this, the existence of a single negative mode
for the $N=1$ instantons supports their decay interpretation.

We would like to stress that the tunneling process between vacua of equal or similar energy densities may not be of solely academic interest: recent studies of eternal inflation have in fact made implicit use of such tunneling processes and have even shown that, under certain circumstances, these similar-height tunneling processes can be the dominant tunneling processes in the multiverse \cite{Johnson:2011aa,Lehners:2012wz}. In an equal-height decay process, the vacuum energy inside of the nucleated bubbles will be identical to that in the starting vacuum, yet other physical quantities will of course be different in general. In a string theoretic context, the value of the scalar field $\varphi$ can determine certain properties, such as coupling constants, of the low energy physics, and thus there will in general be a clear physical distinction between the two vacua. Also, if one considers a more realistic model where other fields are coupled to the tunneling scalar field
one can observe other interesting phenomena like particle creation during tunneling and even more dramatic changes, {\it e.g.} if fermions
have Yukawa couplings to the tunneling field \cite{Rubakov:1984pa,Lavrelashvili:1978zv}. It is interesting to observe that the rates of tunneling in between the two vacua will be the same, regardless of whether one tunnels from the first to the second, or from the second to the first vacuum. This is because the tunneling is mediated via the same instanton solution, and the background subtraction that one performs in calculating the tunneling rate is identical in both cases, as it depends only on the vacuum energy. Thus, such a system will result in a fractal spacetime structure with the overall volume being divided equally between the two vacua in the late future limit.


We conclude with a few notes on future directions: the oscillating instantons which we investigated in present paper exist only due to gravity.
There are no such solutions in (3+1)-dimensional theory in flat spacetime.
On the other hand in scalar field theory there are so-called Fubini instantons \cite{Fubini:1976jm}
describing tunneling without barrier.
Recently gravitating versions of Fubini instantons were investigated in \cite{Lee:2012ug}.
Since in this case the solutions exist in flat as well as in curved spacetime,
it will be interesting to apply a similar analysis to the Fubini instantons
and see what the effects of gravity are in that case.

On a more technical side, we point out that in the present paper we studied a wide class of instantons with the function $Q$ entering the effective fluctuation potential being positive everywhere.
In the situation where $Q$ becomes  negative somewhere along the trajectory,
the perturbation potential becomes singular and the validity of its derivation needs a more careful analysis.
We leave these subtle questions for further investigation.

\section*{Acknowledgements}

\par
This work started during the visit of one of us (G.L.) to the Albert-Einstein-Institute, Potsdam, Germany.
G.L. would like to thank the Quantum Gravity group of this institute and especially Hermann Nicolai for
kind hospitality. L.B. and J.L.L. gratefully acknowledge the support of the European Research Council via the
Starting Grant numbered 256994.
G.L. acknowledges support from Swiss National Science Foundation SCOPES grant 128040.



\newpage

\end{document}